\documentclass[twoside]{article}
\usepackage{amsmath}
\usepackage{amsfonts}
\usepackage{amssymb}
\usepackage{epsf}
\usepackage[dvips]{color}
\font\bfit=cmbxti10 at 10pt

\catcode`\@=11
\long\def\@makefntext#1{
\protect\noindent \hbox to 3.2pt {\footnotesize\hskip-.9pt  
$^{{\@thefnmark}}$\hfil}#1\hfill}                  

\def\thefootnote{\fnsymbol{footnote}}
\def\@makefnmark{\hbox to 0pt{$^{\@thefnmark}$\hss}}        

\def\fnt#1#2{\footnotetext{\kern-.3em
          {$^{\mbox{\scriptsize #1}}$}{#2}}}

\renewcommand{\thefootnote}{\fnsymbol{footnote}}
\newcommand{\alphfootnote}
          {\setcounter{footnote}{0}
           \renewcommand{\thefootnote}{\scriptsize\alph{footnote}}}

\renewcommand{\thefootnote}{\fnsymbol{footnote}}  

\def\ps@myheadings{\let\@mkboth\@gobbletwo
\def\@oddhead{\hbox{}
\rightmark\hfil\footnotesize\thepage}   
\def\@oddfoot{}\def\@evenhead{\footnotesize\thepage\hfil
\leftmark\hbox{}}\def\@evenfoot{}
\def\sectionmark##1{}\def\subsectionmark##1{}}

\def\runninghead#1#2{\pagestyle{myheadings}
\markboth{{\protect\footnotesize\it{\quad #1}}\hfill}
{\hfill{\protect\footnotesize\it{#2\quad}}}}
\def\fpage#1{\begingroup
\voffset=.3in
\thispagestyle{empty}\begin{table}[b]\centerline{\footnotesize #1}
          \end{table}\endgroup}

\oddsidemargin=\evensidemargin
\renewcommand\section{\@startsection {section}{1}{\z@}%
                                  {-10.5pt \@plus -1pt \@minus -.2pt}%
                                  {3.5pt \@plus.2ex}
                                   {\bf}}
\renewcommand\subsection{\@startsection{subsection}{2}{\z@}%
                                    {-8.5pt\@plus -1pt \@minus -.2pt}%
                                    {2.5pt \@plus .2pt}
                                     {\bfit}}
\renewcommand\subsubsection{\@startsection{subsubsection}{3}{\z@}%
                                     {-8.5pt\@plus -1pt \@minus -.2pt}%
                                     {2.5pt \@plus .2pt}%
                                     {\it}}
\newcommand{\nonumsection}[1] {\vspace{12pt}\noindent{\bf #1}
          \par\vspace{5pt}}
\newcounter{appendixc}
\newcounter{subappendixc}[appendixc]
\newcounter{subsubappendixc}[subappendixc]
\renewcommand{\thesubappendixc}{\Alph{appendixc}.\arabic{subappendixc}}
\renewcommand{\thesubsubappendixc}
          {\Alph{appendixc}.\arabic{subappendixc}.\arabic{subsubappendixc}}

\renewcommand{\appendix}[1] {\vspace{12pt}
        \refstepcounter{appendixc}
        \setcounter{figure}{0}
        \setcounter{table}{0}
        \setcounter{lemma}{0}
        \setcounter{theorem}{0}
        \setcounter{corollary}{0}
        \setcounter{definition}{0}
        \setcounter{equation}{0}
        \renewcommand{\thefigure}{\Alph{appendixc}.\arabic{figure}}
        \renewcommand{\thetable}{\Alph{appendixc}.\arabic{table}}
        \renewcommand{\theappendixc}{\Alph{appendixc}}
        \renewcommand{\thelemma}{\Alph{appendixc}.\arabic{lemma}}
        \renewcommand{\thetheorem}{\Alph{appendixc}.\arabic{theorem}}
        \renewcommand{\thedefinition}{\Alph{appendixc}.\arabic{definition}}
        \renewcommand{\thecorollary}{\Alph{appendixc}.\arabic{corollary}}
        \renewcommand{\theequation}{\Alph{appendixc}.\arabic{equation}}
        \noindent{\bf Appendix \theappendixc #1}\par\vspace{5pt}}
\newcommand{\subappendix}[1] {\vspace{12pt}
        \refstepcounter{subappendixc}
        \noindent{\bf Appendix \thesubappendixc. {\kern1pt \bfit #1}}
          \par\vspace{5pt}}
\newcommand{\subsubappendix}[1] {\vspace{12pt}
        \refstepcounter{subsubappendixc}
        \noindent{\rm Appendix \thesubsubappendixc. {\kern1pt \it #1}}
          \par\vspace{5pt}}

\newcommand{\textlineskip}{\baselineskip=13pt}
\newcommand{\smalllineskip}{\baselineskip=10pt}

\newcommand{\copyrightheading}[4]
  {\vspace*{-2.5cm}\smalllineskip{\flushleft
  {\footnotesize Quantum Information and Computation, {Vol.~#1}, {No.~#2}
 {(#3)} {#4}}\\
  {\footnotesize \copyright\kern2pt Rinton Press}\\
    }}

\def\abstracts#1#2#3{{
          \centering{\begin{minipage}{4.5in}\footnotesize\baselineskip=10pt
          \parindent=0pt #1\par 
          \parindent=15pt #2\par
          \parindent=15pt #3
          \end{minipage}}\par}} 

\renewenvironment{thebibliography}[1]
        {\frenchspacing
         \small\rm\baselineskip=11pt
         \begin{list}{\arabic{enumi}.}
        {\usecounter{enumi}\setlength{\parsep}{0pt}     
          \setlength{\leftmargin 12.7pt}{\rightmargin 0pt}
         \setlength{\itemsep}{0pt} \settowidth
          {\labelwidth}{#1.}\sloppy}}{\end{list}}

\newcounter{itemlistc}
\newcounter{romanlistc}
\newcounter{alphlistc}
\newcounter{arabiclistc}

\newcommand{\fcaption}[1]{
        \refstepcounter{figure}
        \setbox\@tempboxa = \hbox{\footnotesize Fig.~\thefigure. #1}
        \ifdim \wd\@tempboxa > 5in
           {\begin{center}
        \parbox{5in}{\footnotesize\smalllineskip Fig.~\thefigure. #1}
            \end{center}}
        \else
             {\begin{center}
             {\footnotesize Fig.~\thefigure. #1}
              \end{center}}
        \fi}

\newcommand{\tcaption}[1]{
        \refstepcounter{table}
        \setbox\@tempboxa = \hbox{\footnotesize Table~\thetable. #1}
        \ifdim \wd\@tempboxa > 5in
           {\begin{center}
        \parbox{5in}{\footnotesize\smalllineskip Table~\thetable. #1}
            \end{center}}
        \else
             {\begin{center}
             {\footnotesize Table~\thetable. #1}
              \end{center}}
        \fi}

\def\pmb#1{\setbox0=\hbox{#1}
          \kern-.025em\copy0\kern-\wd0
          \kern.05em\copy0\kern-\wd0
          \kern-.025em\raise.0433em\box0}

\newcommand{\proof}[1]{{\bf Proof.} #1 $\Box$.}

\def\FigName{figure}%
\newbox\captionbox
\long\def\@makecaption#1#2{%
  \ifx\FigName\@captype
    \vskip\abovecaptionskip
    \setbox\tempbox\hbox{{\figurecaptionfont #1\hskip1em #2}}
          \ifdim\wd\tempbox< 28pc
          \centerline{\box\tempbox}
          \else
          {\figurecaptionfont #1\hskip1em #2\par}
\fi\else
          \setbox\tempbox\hbox{{\tablecaptionfont #1\hskip1em #2}}
          \ifdim\wd\tempbox< 28pc 
          \centerline{\box\tempbox}
          \else
          {\tablecaptionfont #1\hskip1em #2\par}%
          \fi   
 \vskip\belowcaptionskip
 \fi}
\InputIfFileExists{psfig.sty}
{\typeout{^^Jpsfig.sty inputed...ok}}{\typeout{^^JWarning: psfig.sty could be be found.^^J}}
\InputIfFileExists{epsfsafe.tex}
{\typeout{^^Jepsfsafe.tex inputed...ok}}
                              {\typeout{^^JWarning: epsfsafe.tex could not be found.^^J}}
\InputIfFileExists{epsfig.sty}
{\typeout{^^Jepsfig.sty inputed...ok}}{\typeout{^^JWarning: epsfig.sty could not be found.^^J}}
\InputIfFileExists{epsf.sty}
{\typeout{^^Jepsf.sty inputed...ok}}{\typeout{^^JWarning: epsf.sty could not be found.^^J}}%
\def\fps@figure{tbp}
\def\ftype@figure{1}
\def\ext@figure{lof}
\def\fnum@figure{Fig.\ \thefigure}
\def\square{\hbox{${\vcenter{\vbox{                  
   \hrule height 0.4pt\hbox{\vrule width 0.4pt height 6pt
   \kern5pt\vrule width 0.4pt}\hrule height 0.4pt}}}$}}

\def\1#1{{\bf #1}}
\def\2#1{{\cal #1}}
\def\4#1{{\tt #1}}
\def\5#1{{\sf #1}}
\def\6#1{{\frak #1}}
\def\7#1{{\Bbb #1}}
\def\8#1{{\rm #1}}
\def\9#1{{\cal #1}}
\newtheorem{thm}{Theorem}

\newtheorem{lem}[thm]{Lemma}
\newtheorem{prop}[thm]{Proposition}
\newtheorem{cor}[thm]{Corollary}

\def\<{\langle}
\def\>{\rangle}

\def\H#1{L_2(\7F_d^{#1})}
\def\Ew#1#2{ #1 (#2)}
  
\renewcommand{\thefootnote}{\fnsymbol{footnote}}  

\begin{document}
\setlength{\textheight}{8.0truein}    

\runninghead{Cluster states, algotithms and graphs}
            {Dirk Schlingemann}

\normalsize\textlineskip
\thispagestyle{empty}
\setcounter{page}{1}


\vspace*{0.88truein}

\alphfootnote

\fpage{1}

\centerline{\bf CLUSTER STATES, ALGORITHMS AND GRAPHS}
\vspace*{0.035truein}
\centerline{\footnotesize DIRK SCHLINGEMANN}
\vspace*{0.015truein}
\centerline{\footnotesize\it Institut for Mathematical Physics, TU-Braunschweig, Mendelssohnstr. 3}
\baselineskip=10pt
\centerline{\footnotesize\it Braunschweig, 38106, Germany}

\vspace*{0.21truein}

\abstracts{The present paper is concerned with the concept of the one-way quantum computer, beyond binary-systems, and its relation to the concept of stabilizer quantum codes. This relation is exploited to analyze a particular class of quantum algorithms, called graph algorithms, which correspond in the binary case to the Clifford group part of a network and which can efficiently be implemented on a one-way quantum computer. These algorithms can ``completely be solved" in the sense that the manipulation of quantum states in each step can be computed explicitly. Graph algorithms are precisely those which implement encoding schemes for graph codes. Starting from a given initial graph, which represents the underlying resource of multipartite entanglement, each step of the algorithm is related to a explicit transformation on the graph.}{}{}

\vspace*{10pt}


\vspace*{1pt}\textlineskip    
\section{Introduction}
The concept of the {\em one-way quantum computer}, which has first been introduced by Ro\-bert~Raus\-sen\-dorf and Hans~Briegel \cite{BrieRau00,BrieRau01,BrieRau01b,BrieRau01c,BrieBrowRau02,BrieRau02}, describes a realization of of quantum algorithms which goes beyond the usual network picture. A highly entangled multi-partied state, called {\em cluster state} \cite{BrieRau01b}, is the basic resource for running quantum computations. The sequential application of {\em von Neumann measurements} to single qubits in an appropriate order, implements a quantum algorithm. It has been proven that every logical network can be simulated in this way \cite{BrieRau02,BrieBrowRau03}. 

For describing the preparation procedure of a cluster state, we consider finitely many qubits, which are arranged in a cubic lattice $J=\{1,\cdots,n\}^2$. Furthermore, we suppose that an Ising interaction can be switched on and off between neighbored qubits in a controlled manner. For an appropriate description of this ``interaction pattern", it is convenient to introduce the symmetric matrix $\Lambda=(\Lambda(i,j)|i,j\in I)$ whose entries are $\Lambda(i,j)=1$ for neighbored positions $i,j$ and zero else. The preparation of the cluster state can be performed in two main steps. In the first step, a product state is prepared by addressing  each qubit individually in the ``standard state", which can be represented by the vector $\frac{1}{\sqrt{2}}(|0^i\rangle+|1^i\rangle)$. Here $(|0^i\rangle,|1^i\rangle)$ is an orthonormal basis (computational basis) which spans the system Hilbert space $H_i\cong \7C^2$ corresponding to the qubit at position $i\in I$. In the second step, the next-neighbor Ising interaction is switched on for a suitable time interval and then switched off. This produce, which will be explained in more detail later,
creates entanglement between the cluster qubits and yields the desired
{\em cluster state}.

A crucial point is here, that a cluster state can completely be characterized by its {\em stabilizer group}. Namely, a vector $\Psi_{\Lambda}\in H_I:=\otimes_{i\in I}H_i$, which represents the cluster state, fulfills a family of eigenvalue equations
\begin{equation}\label{stab-0}
\prod_{j\in I}\1z_j^{\Lambda(j,i)}\1x_i\Psi_{\Lambda}=\pm \Psi_{\Lambda} \ \ ,
\end{equation}
where each of the equations is labeled by a qubit position $i\in I$. Here $\1x_i$ ($\1z_i$) the are unitary operators on $H_I$, acting by $\sigma_x$- ($\sigma_z$-) Pauli operators at position $i\in I$ and trivially on the remaining positions. The operators $\prod_{j\in I} \1z_j^{\Lambda(j,i)}\1x_i$, $i\in I$, are called ``correlation operators". They commute mutually and therefore generate an abelian group, the stabilizer group of the cluster state. The cluster state can therefore by viewed as a stabilizer code with one-dimensional range.

This shows that cluster states are in fact directly related to stabilizer codes: They {\em are} stabilizer codes. In the usual network picture the ``Clifford part" of an algorithm is composed of CNOT gates, local Hadamard transforms and $\pi/2$-phase gates. This part can be simulated on a one-way quantum computer by just performing measurements in eigenbases of the Pauli operators. An important observation is the following: Suppose one applies a local measurement in one of the eigenbases of the Pauli operators to the cluster state. Furthermore, one selects the state which corresponds to the ``standard measurement outcome $0$" for each measured qubit. Then this state fulfills a set of eigenvalue equations, similar to (\ref{stab-0}), from which follows that the received state is again a stabilizer code with one-dimensional range. Using the fact that each ``stabilizer code" has a representation by a ``graph code" \cite{SchlWer00,Schl02,GraKlaRoe02}, one knows already, that one obtaines a cluster state again (modulo local Hadamard transformations) \cite{BrieBrowRau03}. Thus the Clifford part of an algorithm can be viewed as a sequence of transformations on graphs. These aspects are based on the fact that the theory of stabilizer codes is linked to cluster states by identifying cluster states with ``graph codes". As described above, the symmetric matrix  $\Lambda=(\Lambda(i,j)|i,j\in I)$ can be viewed as an ``interaction pattern". But $\Lambda$ can also be viewed as an ``adjacency matrix" of a graph, where two vertices $i,j$ are connected if $\Lambda(i,j)\not=0$. According to \cite{SchlWer00}, the equations (\ref{stab-0}) determine the graph code associated with $\Lambda$. 

The theory of ``stabilizer codes" is one of the most prominent techniques, for constructing quantum error correcting schemes \cite{CaSho95,CaRainShoSl96a,CaRainShoSl96b,ClGot96,Got96}. For a review, we refere the reader to \cite{Got97}. Based on the results of \cite{Schl02}, the general theory of stabilizer codes from a graph code point of view, is discussed in Section~\ref{stab-codes}. We provide the results which are needed for the analysis of cluster states and local von Neumann measurements. In particular, these results enable an {\em explicit} computation of the stabilizer group of the state which is obtained after ``standard measurements" in eigenbases of the Pauli operators has been applied. 

A general description, how algorithms are implemented on the one-way quantum computer, is given in Section \ref{algos}. We use here the Heisenberg picture in order to stress the operational aspects. The fundamental resource, the cluster state, is determined by an ``interaction graph". Each qudit is labeled by a vertex of the graph. Those qudits which are placed at the ``input vertices" are prepared in the state that carries the initial quantum information. All the remaining qudits, positioned at the ``output vertices" as well as the ``measuring vertices" are prepared individually in the non-binary analogue of the ``standard state". The ``output vertices" label the positions of the qudits which carry the processed information after the computation has been preformed. The ``measuring vertices" label the positions of those qudits to which the local measurement operations are applied. The interaction graph describes one elementary step of the dynamics which is performed after the initial preparation has been completed. This is the essential operation which creates the entanglement between the cluster qudits: The edges corresponding to those pairs of qudits which interact. A computational process is now realized by performing local von Neumann measurements sequentially at the ``measuring vertices". 

For the implementation of an algorithm by a one-way quantum computer we have to solve two main tasks: 

\begin{itemize}
\item
The ``first task" is related to the fact that all measurement outcomes are completely random. In order to perform a deterministic algorithm, we have to ``compensate the randomness" by appropriate conditional local unitary operations. This is analogous to the situation within a teleportation protocol. Here Bob has also to perform a local unitary operation conditioned on the measurement result he received from Alice. Heuristically the process of one-way quantum computing can be  interpreted as a sequence of teleportation sachems which propagate the processed information through the cluster. For a particular measurement result, called the ``standard measurement outcome", we can keep the system as it is, i.e. we do not have to perform a correction by means of a local unitary operation. 

\item
The ``second task" is to find a solution of the following problem: There may be different measurement strategies which lead to the same algorithm. Therefore, we are interested in computing the algorithm explicitly (as a unitary transformation or isometry) from the measurement strategy. Provided the problem of compensating the randomness has been solved,  it remains to analyze how standard measurements operates on a cluster for a given graph.  

\end{itemize}
\begin{figure}[h]
\vspace*{13pt}
\centerline{\epsfig{file=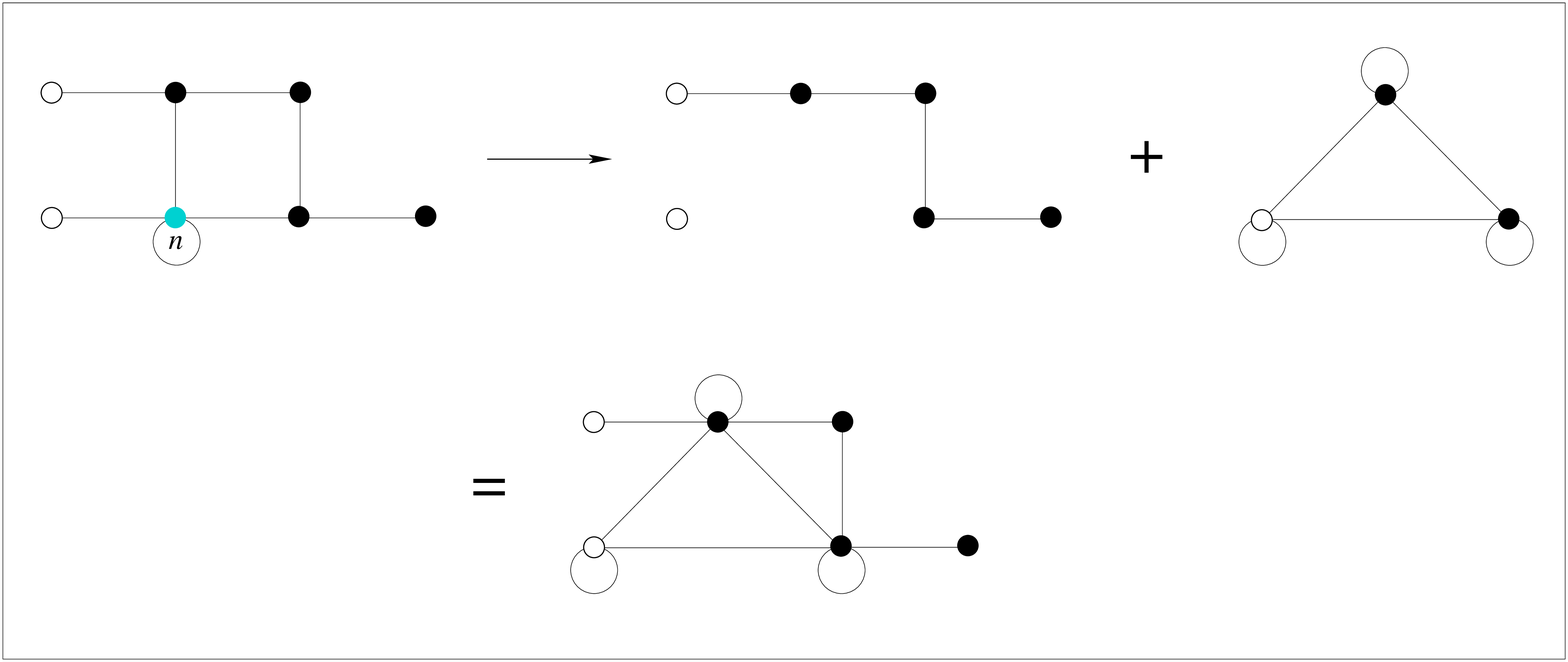, width=13cm}} 
\vspace*{13pt}
\fcaption{Example for removing the measuring vertex $n$ with self-link.}
\label{remove-intro-1}
\end{figure}
\begin{figure}[h]
\vspace*{13pt}
\centerline{\epsfig{file=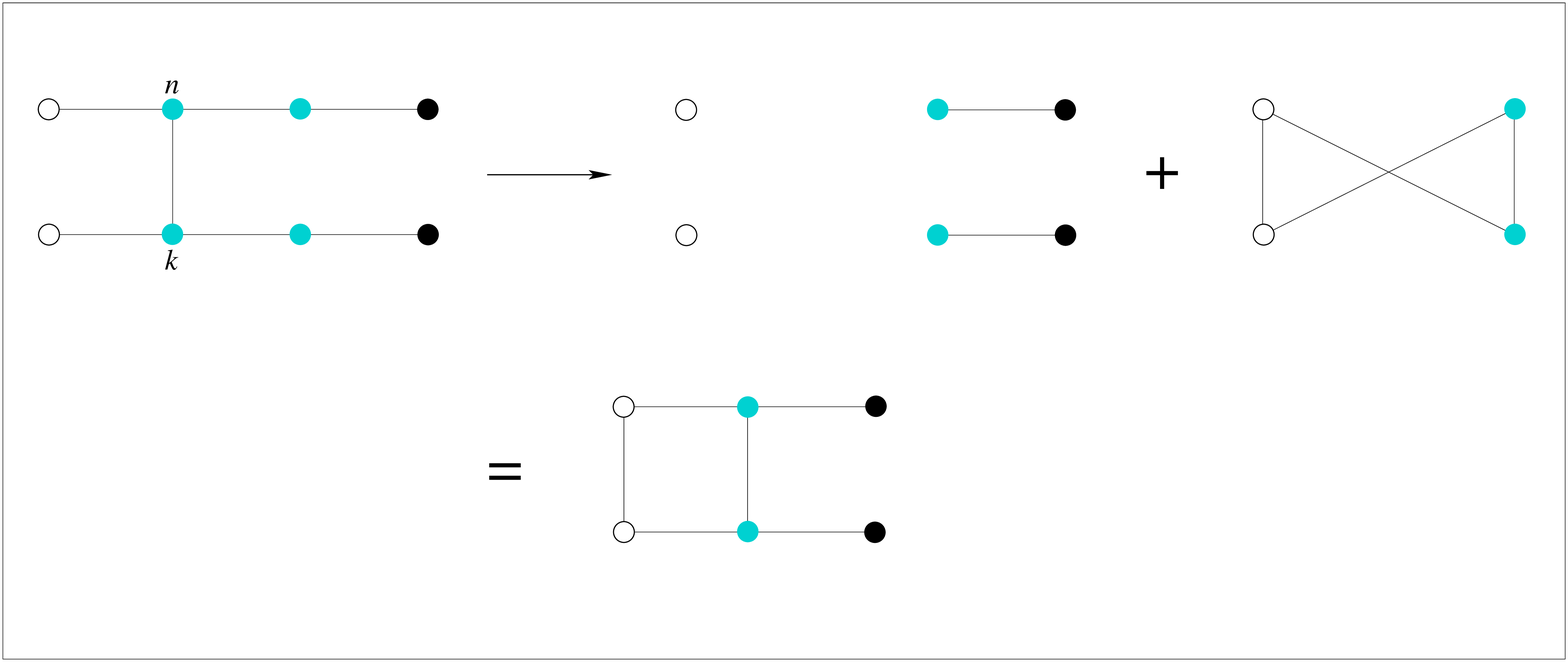, width=13cm}} 
\vspace*{13pt}
\fcaption{Removing two measuring vertices $n$, $k$ that are connected.}
\label{remove-intro-2}
\end{figure}

A special class of algorithms, called ``graph algorithms", is discussed in Section~\ref{graph-algo}. These algorithms are based on ``graph measurements", a particular class of local measurements which are non-binary generalizations of measurements in eigenbases of Pauli operators. For these measurements, both tasks, ``the compensation of randomness" as well as computing the algorithm from the measurement strategy can be solved explicitly. First of all, we show that each graph measurement can be related to an equivalent measurement strategy that only uses measurements in ``$x$-basis" which is the $d$-dimensional analogue of the $\sigma_x$-eigenbasis. In view of the first task, we give an explicite construction of local unitary operations that compensate the randomness of measurements in $x$-basis. At this point we have to assume that the underlying interaction graph is ``basic". This property ensures that sufficient entanglement is created by the dynamics for processing and propagating information through the cluster. The second task is concerned with the computation of standard measurements in $x$-basis which can directly be computed by applying an appropriate transformation to the underlying interaction graph. This transformation ``removes" the corresponding measuring vertices and builds up a new graph on the remaining vertices. For the qubit case, there are simple graphical roles in order to compute this transformation:

\begin{itemize}
\item
Suppose the measuring vertex has a selflink. Then take the subgraph by removing the measuring vertex. Add further edges by mutually connecting those vertices which are linked to the removed vertex including self links. Fig.~\ref{remove-intro-1} and Fig.~\ref{remove-intro-2} represent this procedure.

\item
Suppose now that two measuring, without selflinks, are connected by an edge. Then we can proceed as follows: Build the subgraph by removing both vertices. Then add edges by connecting each vertex, which is linked to one of the removed vertices, with every vertex that is linked to the other one. (See Fig.~\ref{remove-intro-2} for illustation).  
\end{itemize}

Unfortunately, not all sets of measuring vertices can directly be removed. This situation occurs, however, if one is concerned with a measuring vertex, which is not connected to another measuring vertex. In this case we have to apply a ``local Fourier transform" (Hadamard transform) at a neighboring output vertex which ``creates" a new measuring vertex connected to the previous one. Then we can apply the corresponding removing procedure, described above.  

\begin{figure}[h]
\vspace*{13pt}
\centerline{\epsfig{file=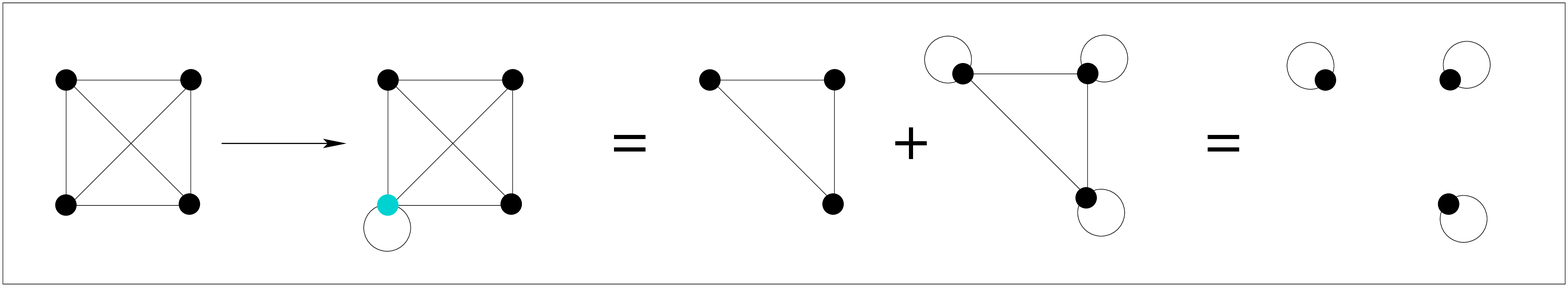, width=13cm}} 
\vspace*{13pt}
\fcaption{The four qubit cluster state for the totally connected graph has persistency $=1$. One just has to perform a measurement in $y$-basis at one of the qubits. Removing the measurement vertex yields a totally disconnected graph corresponding to a product state.}
\label{persist}
\end{figure}

Our techniques provide some additional applications which goes beyond the analysis of algorithms. Examples are the simulation of interactions, quantitative analysis of multipartite entanglemnet, and quantum error correction: 

\begin{itemize}
\item
Standard measurements in $x$-, $y$-, and $z$-bases can be represented as operations on the underlying interaction graph. For a given measurment strategy one obtains a new graph which represents a different interaction. Thus our method provides a tool for computing interactions which can be ``simulated".
 
\item
The persistency of a pure multipartite quantum state is the minimal number of local von Neumann measurements which one has to perform to get a product state \cite{BrieRau01b}. This is a quantity that measures the amount of multipartite entanglement. Our technique to remove measurement vertices can be used to compute bounds on the persistency (See Fig.~\ref{persist}).   

\item
Graphs, which are related to quantum error correcting codes \cite{SchlWer00}, are ``basic" which implies that each graph code can directly be implemented by a graph algorithm. Therefore, by the equivalence of graph and stabilizer codes \cite{Schl02,GraKlaRoe02}, each stabilizer code can be realized by a graph algorithm.  
\end{itemize}

\section{On a general theory of stabilizer codes}
\label{stab-codes}
This section is dedicated to the stabilizer formalism \cite{CaSho95,CaRainShoSl96a,CaRainShoSl96b,Got96,Got97} from a graph code point of view \cite{Schl02}. We supply here useful techniques and results for the analysis of cluster states and local measurement operations performed upon. To setup the formalism, we introduce the concept of discrete phase space as a symplectic space over a finite field and study basic symplectic transformations. By performing the ``canonical quantization procedure", the discrete Weyl algebra results as an appropriate description for the observables of the quantum system. The symplectic transformations can be implemented by unitary operators, yielding the discrete analogue of the ``metaplectic group". The concept of stabilizer algebras (groups), which are particular subalgebras of the Weyl algebra, plays a central role for our analysis. They are naturally represented on the Hilbert space of the system of qudits under consideration. We show here that this representation can explicitly be decomposed into irreducible representations, each occurring with the same multiplicity. Stabilizer codes are precisely defined by the corresponding multiplicity spaces.  

\subsection{The discrete phase space}
The configuration space of a classical digit is an alphabet of $d$ letters, modeled by a $d$-elementary set. As far as our analysis is concerned, we model the finite ``alphabet" by the cyclic fields $\7F_d=\{0,1,\cdots,d-1|\mbox{addition and multiplication modulo d}\}$. For binary systems we are concerned with the field of two elements $\7F_2=\{0,1\}$. A configuration of a classical register is given by a set of {\em positions} $I$ and a tuple $q^I=(q(i)|i\in I)$ in $\7F_d^I$, i.e. a register entry $q(i)$ from the alphabet $\7F_d$ is assigned to each position $i$. In particular, the configuration space of a classical register is a liner space over the finite field $\7F_d$. The corresponding {\em phase space} is modeled by the vector space $\Xi_d^I=\7F_d^I\times\7F_d^I$. The first vector entry of a point $(p^I|q^I)\in\Xi^I_d$ is interpreted as {\em momentum} and the second entry as the {\em position}. The phase space is equipped with a canonical {\em symplectic} form. It assigns to a pair $(p_1^I|q_1^I),(p_1^I|q_1^I)$ of phase space vectors  the ``value" $\langle p_1^I,q_2^I \rangle - \langle p_2^I,q_1^I \rangle$ in $\7F$, with $\langle p^I,q^I\rangle =\sum_{i\in I}p(i)q(i)$. 

Symplectic (canonical) transformations are linear transformations that preserves the symplectic structure. The first type of symplectic transformations, which we consider here, is interpreted as ``discrete dynamics" on phase space. To a symmetric matrix $\Gamma=(\Gamma(i,j)=\Gamma(j,i)|i,j\in I)$ we associate the symplectic transformation that maps a  phase space vector $(p^I|q^I)$ to $(p^I-\Gamma q^I|q^I)$. This ``shear" transformation is viewed as one elementary step of evolution. 
The second type of symplectic transformations, that is relevant for our analysis, is determined by a matrix
$C^J_I=(C(i,j)\in\7F_d|i\in I,j\in J)$ which is invertible. That is, there is a matrix $\bar C_J^I=(\bar C(j,i)|i\in I,j\in J)$ such that $C^J_I\bar C^I_J=1^J_J$ and $\bar C^I_J C^I_J=1^I_I$ holds, where $1^I_I$ is the matrix of the unit operator on $\7F_d^I$. The matrix $C^J_I$ induces a symplectic transformation which maps the vector $(p^I|q^I)$ to $(C^J_Iq^I|-{^t\bar C}^J_Ip^I)$\footnote{We denote by ${^tM}^J_I$ the {\em transpose} of a matrix $M_J^I$.}.

\subsection{Discrete Weyl systems}
The Hilbert space, describing a quantum register of qudits, is the space $\H{I}$ of complex valued functions on $\7F_d^I$ and its complex dimension is $d^{|I|}$ where $|I|$ is the number of elements in $I$. The scalar product of two functions $\psi_1,\psi_2$ is given by
\begin{equation}
\langle\psi_1,\psi_2\rangle=\int\8dq^I\ \bar\psi_1(q^I)\psi_2(q^I)
\end{equation}
where the integration is performed with respect to the normalized Haar measure of the additive group $\7F_d^I$, i.e. a function $f$ on $\7F_d^I$ is integrated according to
\begin{equation}
\int \8dq^I \ f(q^I)=d^{-|I|}\sum_{q^I\in\7F_d^I} f(q^I) \; .
\end{equation}

The {\em shift operator} $\1x(q^I)$, associated with position vector $q^I$, is the unitary operator, which translates (shifts) a function $\psi\in\H{I}$ by $q^I$. The value of the shifted function at $q^I_1$ is $
(\1x(q^I)\psi)(q_1^I)=\psi(q_1^I-q^I)$. The {\em multiplier operator} $\1z(p^I)$, which is associated with the momentum $p^I$, is the unitary multiplication operator, acting on a function $\psi\in\H{I}$ by
\begin{equation}
(\1z(p^I)\psi)(q^I)=\chi(p^I|q^I)\psi(q^I)\; \text{ with } \;
\chi(p^I|q^I)=\8e^{\frac{2\pi\8i}{d}\langle p^I, q^I\rangle} \; .
\end{equation}
By combining shift and multiplier operators, we associate to each point in phase space $(p^I|q^I)\in\Xi^I$ the unitary operator
\begin{equation}
\1w(p^I|q^I)=\1z(p^I)\1x(q^I)\; ,
\end{equation}
which we call the {\em Weyl operator} associated with $(p^I|q^I)$. They satisfy a discrete version of the Weyl (or canonical) commutation relations
\begin{equation}\label{discr-ccr}
\1w(p_1^I|q_1^I)\1w(p_2^I|q_2^I)=\chi(p_2^I|q_1^I)\1w(p^I_1+p_2^I|q_1^I+q_2^I)\; .
\end{equation}
The product of two Weyl operators is again a multiple of a Weyl operator. By taking all their linear combinations, we obtain an algebra $\6A(I)$, the discrete Weyl algebra, of operators which act on the Hilbert space $\H{I}$. This algebra serves as the observable algebra of a quantum register. For a subset $J\subset I$ of positions, the Weyl operators $\1w(p^J|q^J)$ generate sub-algebra $\6A(J)\subset\6A(I)$, identifying a quantum sub-register by operating only non-trivially on the positions $J$\footnote{The Weyl operators form a basis for the space of all linear operators on $\H{I}$. Hence the algebra $\6A(I)$ is the full matrix algebra of all linear operators acting on $\H{I}$.}.

The states of the quantum register, with qudit positions $I$, are given by the density operators $\rho$ on $\H{I}$, i.e. the expectation value of an operator $A$ is evaluated by $\Ew{\omega}{A}=\8{tr}(\rho A)$. One example, the so called {\em standard state} on $\6A(I)$, is the pure state which is implemented by the projection onto the normalized vector $\xi_{[I]}={\xi}_{[0^I]}\in\H{I}$. This vector is given by the function that assigns to each classical register configuration $q^I$ the constant value $1$. A crucial prperty of the vector ${\xi}_{[I]}$ is that it is invariant under all shifts $\1x(q^I)$ and that it is a product vector ${\xi}_{[I]}=\otimes_{i\in I}{\xi}_{[i]}$. By the application of multiplier operators $\1z(p^I)$ to ${\xi}_{[I]}$, we obtain an orthonormal basis, called the {\em $x$-basis},
\begin{equation}
\xi=\bigl\{{\xi}_{[p^I]}:=\1z(p^I){\xi}_{[I]}\bigm |p^I\in\7F^I\bigr\} \; .
\end{equation}
It is a joint eigenbasis for all shift operators $\1x(q^I)$, according to the Weyl commutation relations,
\begin{equation}\label{x-basis}
\1x(q^I){\xi}_{[p^I]}=\chi(p^I|q^I){\xi}_{[p^I]} 
\end{equation}
which immediately implies that the vector ${\xi}_{[p^I]}$ can be viewed as an {\em eigenvector of the momentum operator} subject to the {\em spectral value} $p^I$. 

Furthermore, the {\em $z$-basis} is of importance here. It can be obtained by an application of shift operators to the normalized wave function ${\zeta}_{[I]}$ which is given by 
\begin{equation}
{\zeta}_{[I]}(q^I)=\sqrt{d}^{|I|}\delta(q^I) \; .
\end{equation}
Since ${\zeta}_{[I]}$ is only supported at the zero configuration $q^I=0^I$, it is invariant under all multiplier operators $\1z(p^I)$. As for the standard state, it analogously implements a state which is now invariant under all multiplier operators $\1z(p^I)$. By acting with the shift operators $\1x(q^I)$ on ${\zeta}_{[I]}$, we obtain the $z$-basis
\begin{equation}
\zeta=\bigl\{{\zeta}_{[q^I]}:=\1x(q^I){\zeta}_{[I]}\bigm |q^I\in\7F^I\bigr\} \; .
\end{equation}
Once again, it follows from the Weyl commutation relations (\ref{discr-ccr}) that it is a joint eigenbasis for all multiplier operators $\1z(q^I)$.

\subsection{Implementation of symplectic transformations}
In the classical context, a symmetric matrix $\Gamma$ implements a symplectic transformation on the discrete phase space $\Xi^I$ which we interpret as dynamics. The quantized version of this dynamics is given by the unitary multiplication operator
\begin{equation}\label{defi-dyn}
({u}(\Gamma)\psi)(q^J)=\tau(\Gamma|q^I)\psi(q^I)\;
\text{ with }\; \tau(\Gamma|q^I)=\8e^{\frac{\pi\8i}{d}\langle
q^I,\Gamma q^I\rangle} \; .
\end{equation}
Indeed, ${u}(\Gamma)$ {\em implements} the symplectic transformation $(p^I|q^I)\mapsto(p^I-\Gamma^I_Iq^I|q^I)$ i.e. the commutation relation
\begin{eqnarray}\label{com-rel-0}
{u}(\Gamma)\1w(p^I|q^I)=\tau(-\Gamma|q^I)\1w(p^I-\Gamma q^I|q^I){u}(\Gamma)
\end{eqnarray}
holds. The symmetric matrix $\Gamma$ describes a ``pattern" of two-qudit interactions. This can be visualized by a {\em weighted graph} whose vertices are the positions $i\in I$. Two vertices $i,j$ are connected by an edge if the matrix element $\Gamma(i,j)\not=0$ is non-vanishing. The value of the matrix element $\Gamma(i,j)$ is interpreted as the ``strength" of the interaction. 

The symplectic transformation $(p^I|q^I)\mapsto(C^J_Iq^I|-{^t\bar C}^J_Ip^I)$ is also implemented by a unitary transformation $F_{[C^J_I]}$. It is called the {\em Fourier transform} associated with the invertible matrix $C^J_I$. It identifies the Hilbert space $\H{I}$ with $\H{J}$ according to
\begin{equation}
(F_{[C^J_I]}\psi)(q^J)=\sqrt{d}^{|I|}\int\8dq^I \chi(C^J_Iq^I|q^J)\psi(q^I) \; .
\end{equation}
By construction, the commutation relation
\begin{eqnarray}
F_{[C^J_I]}\1w(p^I|q^I)=\1w(C^J_Iq^I|-{^t\bar C}^J_Ip^I)F_{[C^J_I]}
\end{eqnarray}
follows. For our purpose, we consider those Fourier transforms which are associated with an invertible matrix $C^J_I$, where $I$ and $J$ are disjoint. We also visualize such a Fourier transformation by a weighted graph on the union $I J$:\footnote{Notation and conventions: For the union of two finite sets $K$ and $L$ we often drop the union-symbol: $KL:=K\cup L$.} A vertex $i\in I$ is connected with a vertex $j\in J$ if $C(j,i)\not=0$. The symmetric matrix $C^J_I+{^t C}^I_J$ is the adjacency matrix of the graph. If each vertex of $I$ is connected with precisely one vertex in $J$, then we call $C^J_I$ a {\em connecting matrix} or {\em connecting graph} to put emphases on the graph-theoretical aspects. The corresponding Fourier transforms play a crucial role. They are {\em local unitary} transformations on the qudit systems labeled by $I$. In fact the Fourier transform, associated with a connecting graph, is a tensor product
\begin{equation}
F_{[C^J_I]}=\bigotimes_{i\in I}F_{[C^{j(i)}_i]}
\end{equation}
of Fourier transforms operating independently on each single digit. Namely, a connecting graph $C^J_I$ decomposes into connected components 
\begin{equation}
C^J_I=\sum_{i\in I}C^{j(i)}_i
\end{equation}
where $C^{j(i)}_i$ is the connecting graph consisting of the line with weight $C(j,i)$ connecting the vertex $i\in I$ with its unique counterpart $j(i)$ in $J$.

\subsection{Stabilizers}
An {\em isotropic subspace} of a discrete phase space is characterized by the vanishing of the symplectic form for all pairs of its vectors. Making use of the results of \cite{Schl02}, an isotropic space can be parameterized by a weighted graph $\Lambda$ on a union of four sets the {\em input vertices} $I$, {\em output vertices} $J$, {\em auxiliary vertices} $K$,  and {\em syndrome vertices} $L$. The isotropic space, associated with $\Lambda$, is the subspace which is defined according to\footnote{Notation and conventions: For a matrix $\Theta^N_M$ and for two subsets $K\subset M$, $L\subset N$, we write $\Theta^L_K=(\Theta(l,k)|l\in L,k\in K)$ for the corresponding sub-block.}
\begin{equation}\label{iso}
S_{[I,J,K|\Lambda]}:=\bigl\{(\Lambda^{J}_{J K}q^{J K}|q^{J})\bigm |
\Lambda^{I K}_{J K}q^{J K}=0\bigr\} \; .
\end{equation}
According to the Weyl commutation relations the set of Weyl operators
\begin{equation}\label{weyl-set}
\bigl\{\1w(\Lambda^{J}_{J K}q^{J K}|q^{J})\bigm|q^{J K}\in\8{ker}\Lambda^{I K}_{J K}\bigr\} \; .
\end{equation}
generate an abelian sub-algebra  $\6A(I,J,K|\Lambda)\subset\6A(J)$ which we call the {\em stabilizer algebra} of the isotropic subspace $S_{[I,J,K|\Lambda]}$. The {\em stabilizer group} is just the unitary group generated by (\ref{weyl-set}). The definition of the isotropic space as well as the construction of the stabilizer algebra only depends on the sub-graph $\Lambda^{I J K}_{I J K}$ subject to the complement of the syndrome vertices $I J K$. Despite of this, the edges that connect syndrome vertices with others play an important role for the representation theory of the stabilizer algebra. 

However (see \cite{Schl02}), for parameterizing an arbitrary isotropic space we can restrict ourselves to weighted graphs which we call {\em admissible}. They fulfills the following list of conditions:

\begin{description}
\item{\rm(G1)}
The set of vertices satisfy $|I|+|L|=|J|$.

\item{\rm(G2)}
The submatrix $\Lambda^{J K}_{I K L}$ is invertible.

\item{\rm(G3)} There are no edges that connect syndrome vertices, i.e. $\Lambda^L_L=0$.
\end{description}

\begin{figure}[h]
\vspace*{13pt}
\centerline{\epsfig{file=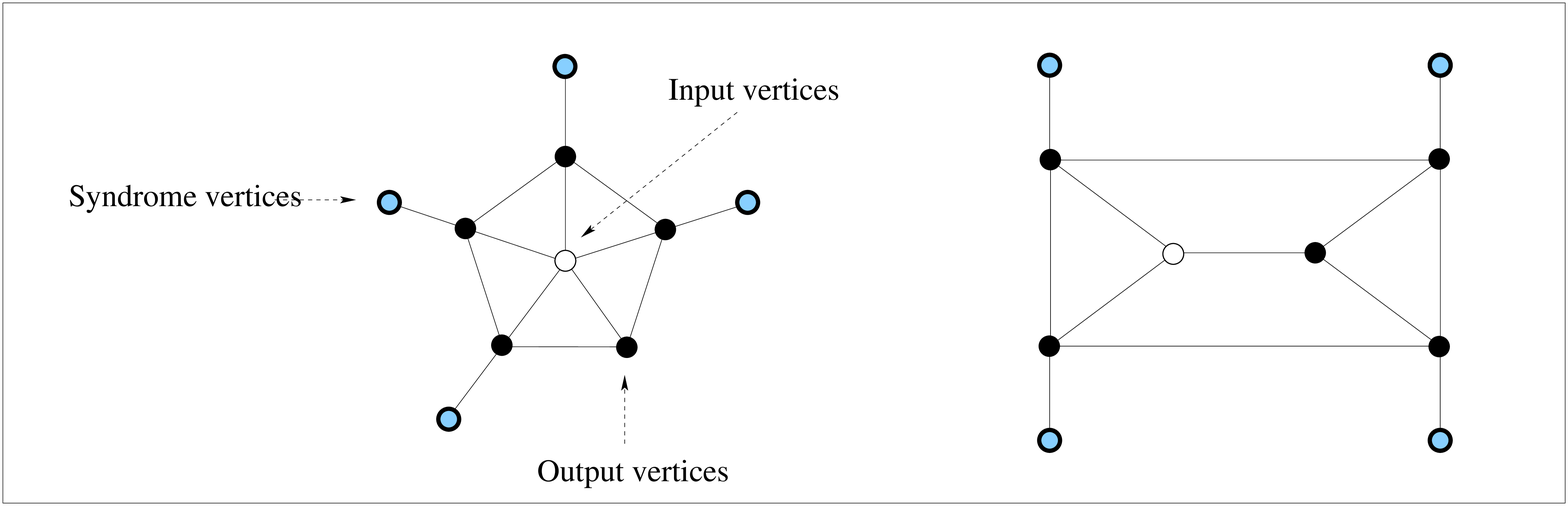, width=13cm}} 
\vspace*{13pt}
\fcaption{Examples for graphs, corresponding to $[[5,1,3]]_d$ stabilizer codes, fulfilling the conditions (G1)-(G3).}
\label{example-0}
\end{figure}

Simple examples for admissible graphs are shown in Fig.~\ref{example-0}. As shown in \cite{SchlWer00}, these graphs correspond to $[[5,1,3]]_d$ stabilizer codes. However, one only has to consider the syndrome vertices if one is interested in an explicit construction of the decoding operation. The encoding itself just makes use of the subgraph which is obtained by wiping out the syndrome vertices.    

Since $\6A(I,J,K|\Lambda)$ is an abelian C*-algebra its representation on $\H{J}$, which we call its {\em natural representation}, can be decomposed into irreducible representations, called {\em characters}\footnote{An irreducible $\varpi$ representation of an abelian C*-algebra $\6C$ is one-dimensional, i.e. it can also be viewed as a state $\varpi:A\mapsto\Ew{\varpi}{A}$ on $\6C$. The functional $\varpi$ is multiplicative and therefore a pure state.}. Each character occurs with a certain multiplicity, i.e. the Hilbert space $\H{J}$ decomposes into a direct sum of multiplicity spaces. The decomposition of the natural representation into irreducibles can directly be derived from the structure of the weighted graph $\Lambda$. In particular the syndrome vertices are used to arrange the combinatorics for obtaining an explicit decomposition. How this can be done, will be discussed in the next section in detail.

\subsection{Representation theory}
The stabilizer algebra $\6A(I,J,K|\Lambda)$ is generated by the set of Weyl operators (\ref{weyl-set}). They form a linearly independent family of operators and since the product of two of them is a multiple of another Weyl operator within this family, it is a basis for  $\6A(I,J,K|\Lambda)$. The condition (G2) implies that the kernel of $\Lambda^{I K}_{J K}$ contains $d^{|L|}$ elements and the dimension of $\6A(I,J,K|\Lambda)$ is therefore $d^{|L|}$. It turns out that for all characters of the stabilizer algebra the
multiplicity spaces have the same dimension. And since the dimension of $\H{J}$ is $d^{|J|}$ it follows from (G1) that the each multiplicity space is $d^{|I|}$ dimensional and therefore isomorphic to $\H{I}$. There are $d^{|L|}$ different characters of the stabilizer algebra and we may label each multiplicity space by a classical register configuration $q^L$ in $\7F^L_d$. The basic idea to perform the decomposition is to construct for each classical register configuration $q^L$ an isometry from $\H{I}$ into $\H{J}$ whose range is precisely the multiplicity space labeled by $q^L$. This is nothing else but a coding operation. The Hilbert space $\H{I}$ is the {\em input space} corresponding to the system we wish to encode, and $\H{J}$ is the {\em output space} in which quantum information is encoded. This point of view is the reason for calling the set $I$ input vertices and the set $J$ output vertices. 

The $x$-basis vectors ${\xi}_{[p^N]}$ and its $z$-basis counterparts ${\zeta}_{[q^N]}=|q^N\rangle$ induce isometric mappings
\begin{eqnarray}\label{iso-stand}
&&\Phi_{[p^N]}:\2H\to \H{N}\otimes\2H, \; \psi\mapsto{\xi}_{[p^N]}\otimes\psi
\\
&&\Pi_{[q^N]}:\2H\to \H{N}\otimes\2H, \; \psi\mapsto{\zeta}_{[q^N]}\otimes\psi 
\end{eqnarray}
were we write $\Phi_{[N]}=\Phi_{[0^N]}$ and $\Pi_{[N]}=\Pi_{[0^N]}$. From these isometries and the unitary ${u}(\Lambda)$ which implements the dynamics for the ``interaction pattern" $\Lambda$, we build the linear operator
\begin{equation}\label{encode}
\1v_{[\Lambda|q^L]}
:=\sqrt{d}^{|JK|}\Phi_{[{I K L}]}^*{u}(\Lambda)\Phi_{[{J K}]}\Pi_{[q^L]}
\end{equation}
that serves as a candidate for identifying the input space $\H{I}$ with a multiplicity space. One also obtains a more explicit expression for the operator $\1v_{[\Lambda|q^L]}$ by applying it to a vector ${\zeta}_{[q^I]}=|q^I\rangle$ of the $z$-basis for the input space and expanding the result in terms of the $z$-basis of the output space
\begin{equation}\label{graph-code}
\1v_{[\Lambda|q^L]}|q^I\rangle
=\sqrt{d}^{2|JK|}\int\8dq^{J K} \
\tau(\Lambda|q^{I J K L}) |q^J\rangle \; .
\end{equation}
Let us suppose that the graph $\Lambda$ has no edges that connects a vertex with itself, then the phases $\tau(\Lambda|q^{I J K L})$ can be obtained by just taking the product over all edges of the graph $\Lambda$, where to each edge $\{i,j\}$ the phase $\exp(\frac{2\pi\8i}{d}q(i)\Lambda(i,j)q(j))$ is assigned. Comparing the expression (\ref{graph-code}) with the construction of graph-codes, outlined in \cite{SchlWer00,Schl02}, we see that $\1v_{[\Lambda|q^L]}$ is nothing else but the encoding isometry of a graph-code whose range is indeed a multiplicity space of a character
of the stabilizer algebra $\6A(I,J,K|\Lambda)$.

The following central theorem states that these intuitions are correct and it shows how the decomposition of the natural representation of a stabilizer algebra is encoded by its associated weighted graph.

\begin{thm}\label{thm-decompose}
Let $\Lambda$ be an admissible graph on the set of vertices $I J K L$. Then the natural representation of the stabilizer algebra $\6A(I,J,K|\Lambda)$ decomposes into a direct sum of characters $\{\varpi_{[q^L]}|q^L\in\7F^L_d\}$. The character $\varpi_{[q^L]}$ is uniquely determined by the Weyl expectation values
\begin{equation}
\Ew{\varpi_{[q^L]}}{\1w(\Lambda^{J}_{J K}q^{J K}|q^J)}=
\tau(\Lambda|q^{J K L})\; , \; \text{with $q^{J K}\in\8{ker}\Lambda^{I K}_{J K}$.}
\end{equation}
The operator $\1v_{[\Lambda|q^L]}$ is an isometry whose range is the multiplicity space of the character $\varpi_{[q^L]}$.
\end{thm}
\proof{
A proof of the theorem can mainly be derived by using the techniques which has been developed in \cite{SchlWer00,Schl02,Schl02b}. To keep the paper self contained, we prove some useful technical lemmas in the appendix. First of all, Lemma \ref{tech-lem-1} ensures that the operators $\1v_{[\Lambda|q^L]}$ are isometric. Lemma \ref{tech-lem-2} states that the range of the isometry $\1v_{[\Lambda|q^L]}$ is in fact the multiplicity space of the character $\varpi_{[q^L]}$ of the stabilizer algebra $\6A(I,J,K|\Lambda)$. To conclude the prove of the theorem, we apply Lemma \ref{tech-lem-3} from which follows that the ranges of the isometries $\1v_{[\Lambda|q^L]}$, $q^L\in\7F_d^L$, are mutually orthogonal and that they ``add up" (as a direct sum) to the full Hilbert space $\H{J}$}  

As an abelian algebra of dimension $d^{|L|}$, $\6A(I,J,K|\Lambda)$ is isomorphic to the abelian algebra of complex function on the ``syndrome configurations" $\7F_d^L$. Due to Theorem \ref{thm-decompose} we can explicitly identify operators in $\6A(I,J,K|\Lambda)$ with functions on $\7F_d^L$. Namely, a Weyl operator $\1w(\Lambda^{J}_{J K}q^{J
K}|q^J)$, $q^{J K}\in\8{ker}\Lambda^{I K}_{J K}$, is just identified with the function $q^L\mapsto\tau(\Lambda|q^{J K L})$. Furthermore, Theorem \ref{thm-decompose} implies that every vector $\psi$ in the range of the isometry $\1v_{[\Lambda|q^L]}$ fulfills the eigenvalue equation
\begin{equation}\label{stab-ew}
\1w(\Lambda^{J}_{J K}q^{J K}|q^{J}) \psi
=\tau(\Lambda|q^{J K L})\psi
\end{equation}
whenever the classical register configuration $q^{J K}$ satisfies the constraint $\Lambda_{J K}^{I K}q^{J K}=0$. This is just what corresponds to the eigenvalue equation (\ref{stab-0}) for a qubit cluster state and there are two special situations of particular interest:

\paragraph*{\it No input vertices:}
The first special case is concerned with admissible graphs for which the set of input vertices is empty $I=\emptyset$. In this case, the vector
\begin{equation}
\Psi_{[\Lambda|q^L]}:=\sqrt{d}^{|JK|}\Phi_{[{K L}]}^*{u}(\Lambda)({\xi}_{[{J K}]}\otimes{\zeta}_{[q^L]})
\end{equation}
is the unique normalized vector in the one-dimensional range of $\1v_{[\Lambda|q^L]}$. The stabilizer algebra $\6A(J,K|\Lambda)=\6A(\emptyset,J,K|\Lambda)$ is generated by Weyl operators which correspond to the isotropic subspace of phase space vectors $(\Lambda^{J}_{J K}q^{J K}|q^{J})$, $q^{J K}\in\8{ker}\Lambda^K_{J K}$.
An immediate consequence of Theorem \ref{thm-decompose} is then the corollary:

\begin{cor}\label{cor-ew}
The family of vectors
$\{\Psi_{[\Lambda|q^L]}|q^L\in\7F_d^L\}$ is an orthonormal basis in $\H{J}$ and each basis vector $\Psi_{[\Lambda|q^L]}$ fulfills the eigenvalue equations
\begin{equation}
\1w(\Lambda^{J}_{J K}q^{J K}|q^{J})\Psi_{[\Lambda|q^L]}=
\tau(\Lambda|q^{J K  L})\Psi_{[\Lambda|q^L]}
\end{equation}
for all classical register configurations $q^{J K}\in\8{ker}\Lambda^K_{J K}$.
\end{cor}

\paragraph*{\it No input and auxiliary vertices:}
The second situation, being concerned here, is even more special. Namely, we now consider admissible graphs with no input and auxiliary vertices $I=K=\emptyset$. Since there is no edge that connects a vertex in $J L$ with a vertex in $\emptyset$, the kernel of $\Lambda^{K=\emptyset}_{J}=0$ is just the full vector space $\7F^J_d$. The corresponding stabilizer algebra $\6A(J|\Lambda):=\6A(\emptyset,J,\emptyset|\Lambda)$ is generated by the Weyl operators $\1w(\Lambda^{J}_{J}q^{J}|q^{J})$ where $q^J$ can be any classical register configuration. Hence, the vectors of the orthonormal basis $\{\Psi_{[\Lambda|q^L]}|q^L\in\7F_d^L\}$ fulfill the eigenvalue equations
\begin{equation}\label{ew-special-1}
\1w(\Lambda^{J}_{J}q^{J}|q^{J})\Psi_{[\Lambda|q^L]}
=\tau(\Lambda|q^{J L})\Psi_{[\Lambda|q^L]}
\end{equation}
for all $q^J$.

\section{One-way quantum computing}
\label{algos}
The results on stabilizer codes, discussed in the previous section, are now used as a basis for investigating the concept of the one-way quantum computer. The algorithms under consideration are given in terms of {\em elementary operations} which are preparation of individual qudits, performing one elementary steps of dynamics, applying local von Neumann measurements and performing local unitary operations, conditioned on measurement results.

\subsection{Elementary operations}
\label{elementary-operations}
For the implementation of algorithms, we discuss here those operations which are regarded as elementary. Namely, local preparations of individual qudits, elemetary steps of a global dynamics, local measurements and conditional local unitary operations.

\paragraph*{\it Local preparation of individual qudits:}
The input qudits at position $I$ are prepared in the input state, carrying the quantum information one wishes to process. The qudits at the remaining positions $N$ are individually prepared in the standard state. In the Heisenberg picture, this operation is described by an encoding channel $E_{I}$, which maps an observable $A\in \6A(I N)$ to the operator 
\begin{equation}
E_{I}(A):=\Phi_{[{N}]}^*A\Phi_{[{N}]}\in\6A(I) \; .
\end{equation}
If the set $I=\emptyset$ is empty, then the operation $E_{\emptyset}$ just prepares the standard state which is implemented by the vector ${\xi}_{[N]}$. To point out the analogy to the binary case we expand it in terms of the $z$-basis, where we sometimes write the basis vector ${\zeta}_{[q^N]}$ as a ``ket" $|q^N\rangle$:
\begin{equation}
{\xi}_{[N]}=\sqrt{d}^{|N|}\int \8dq^N \ |q^N\rangle=\bigotimes_{n\in N}\, \frac{1}{\sqrt{d}}\sum_{q(n)\in\7F_d} \ |q(n)\rangle \; .
\end{equation}
For $d=2$, we indeed obtain a tensor product of the vectors $\frac{1}{\sqrt{2}}(|0^n\rangle+|1^n\rangle)$ over $N$.

\paragraph*{\it Elementary step of discrete dynamics:}
One elementary step of the discrete dynamics, is determined by a weighted graph $\Gamma$ on a set of vertices $N$. It is given by an automorphism 
\begin{equation}
\alpha_\Gamma:\6A(N)\to\6A(N)
\end{equation}
that maps an operator $A\in\6A(N)$ to  
\begin{equation}
\alpha_{\Gamma}(A)={u}(\Gamma)^*A{u}(\Gamma) \; .
\end{equation}
The dynamics operates parallel on all qudits in the cluster within one step and is, in comparison to the usual network picture, a very fast operation. As already mentioned, the graph $\Gamma$ describes the type of two qudit interactions: Two qudits interact if their positions are connected by an edge. In view of this $\Gamma$ is called an ``interaction graph".

\paragraph*{\it Local von Neumann measurements:}
A von Neumann measurement, acting on the qudits at positions $M$, is determined by a choice of a orthonormal basis $\beta=\{\beta_{[q^M]}|q^M\in\7F^M\}$ of $L_2(\7F_d^M)$. In the Heisenberg picture, this operation maps a function $f$ of the classical observable algebra $\6C(M)$ of functions on $\7F_d^M$ to the operator
\begin{equation}\label{vnm}
P_{[\beta,M]}(f):=\sum_{q^M\in\7F_d^M} f(q^M) \ P_{[\beta|q^M]} \ \in\6A(M)
\end{equation}
where $P_{[\beta|q^M]}$ is the projection onto the basis vector $\beta_{[q^M]}$. Von Neumann measurements are mathematically characterized by the property that they preserve the multiplicative structure, in other words, $P=P_{[\beta,M]}$ is an C*-algebra homomorphism. This means that the product $f g$ of two functions is mapped to the product of operators $P(f)P(g)$ and the complex conjugate function $\bar f$ is mapped to $P(\bar f)=P(f)^*$ to the adjoint of $P(f)$. A quantum state $\omega$ on $\6A(M)$ is mapped to the classical probability distribution. The probability for receiving the measurement outcome $q^M$ is then given by the expectation value $\Ew{\omega}{P_{[\beta|q^M]}}$. The values for the classical measurement outcomes under consideration are classical register configurations. In the binary situation one receives, for example, a bit string of length $|M|$. The additive structure of register configurations, allows to choose the zero-register configuration $q^M=0^M$ as a natural ``standard" for the ``initialization" of a classical register. In view of this, we call $q^M=0^M$ the {\em standard measurement outcome} and the corresponding selective measurement operation is called {\em standard measurement operation}. A {\em local von Neumann measurement} is related to a product basis $\beta$, i.e. each vector of the basis is a product vector. The corresponding local measurement operation decomposes into a tensor product $P_{[\beta,M]}=\otimes_{m\in M}P_{[\beta,m]}$, addressing each qudit individually. The $x$-basis $\xi=\{\xi_{[p^M]}|p^M\in\7F^M\}$ as well as the $z$-basis $\zeta=\{\zeta_{[q^M]}|q^M\in\7F^M\}$ consists of product vectors. Thus they correspond to local von Neumann measurements $P_{[\xi,M]}$ and $P_{[\zeta,M]}$, respectively. Since the $x$-basis consists of eigenvectors for all shifts, the measurements in $x$-basis are invariant under shift operations, i.e.
\begin{eqnarray}
\1x(q^M) \ P_{[\xi,M]}(f) \ \1x(q^M)^*=P_{[\xi,M]}(f)
\end{eqnarray}
is valid for all $q^M$ and for all functions $f\in\6C(M)$. Analogously, the measurements in $z$-basis are invariant under all multiplier operations.  Measurements in $x$- and $z$-bases are mostly applied to a subsystem $\6A(M)\subset\6A(I M)$. In this case, the measurement operation $P_{[\xi,M]}$ only operate on the positions $M$ and leave the rest of the system alone. Keeping in mind that for all operators $A\in\6A(I)$ the identity $\Phi_{[p^M]} A\Phi_{[p^M]}^*=A\otimes P_{[\xi|p^M]}$ holds, it is a straight forward observation that the co-isometries $\{\Phi_{[p^M]}^*|p^M\in\7F^M\}$ are Kraus operators for the local measurement operation $P_{[\xi,M]}$, where $\Phi_{[p^M]}^*$ corresponds to the random measurement outcome $p^M$. Analogously, if one is concerned with a local measurement operation in $z$-basis, then the Kraus operators  are given by the co-isometries $\{\Pi_{[q^M]}^*|q^M\in\7F^M\}$. 

\paragraph*{\it Conditional local unitary operations:}
Let $M$ be a set of positions at which local von Neumann measurements have been applied and let $J$ be the set of positions of the remaining qudits. A conditional local unitary operation is given by a channel 
\begin{equation}
C:\6A(J)\to\6C(M)\otimes \6A(J)
\end{equation}
which is implemented by a family of local unitary operators $\{\1s_{[C|p^M]}|p^M\}$
in $\6A(J)$ according to 
\begin{equation}\label{cond-unitary}
[C(A)](p^M)=\1s_{[C|p^{M}]}^* A \1s_{[C|p^{M}]}\; .
\end{equation}
Indeed, this is nothing else but performing the applying local unitary $\1s_{[C|p^M]}$ in case the measurement result $p^M$ was received.   

\subsection{Algorithms}
\label{algorithms}
An algorithm is implemented by applying a sequence of elementary operations, as described above, in an appropriate order. Each algorithm is mathematically modeled by a channel, i.e. completely positive unit-preserving map,
\begin{equation}
T:\6A(J)\to\6A(I)
\end{equation}
that maps an observable of the output system $\6A(J)$ to an observable of the input system $\6A(I)$, i.e. concerning the Schr\"odinger picture, a state of the input system is mapped to a state of the output system. Using the concept of the one-way quantum computer, the channel $T$ is realized by the following procedures:

\smallskip
\noindent
{\bf Step $\>00\>$:}
In the first place, the qudits with positions at the {\em input vertices} $I$ are prepared in the state that carries the quantum information we which to process. The qudits at the {\em output vertices} $J$ and the {\em measuring vertices} $K$ are individually prepared in the standard state. This elementary operation is given by the channel
\begin{equation}\nonumber
E_I:\6A(I J K)\to\6A(I) \; .  
\end{equation}

\smallskip
\noindent
{\bf Step $\>01\>$:}
The next elementary operation is an application of one elementary step of the discrete dynamics 
\begin{equation}\nonumber
\alpha_\Gamma:\6A(I J K)\to \6A(I J K)
\end{equation}
associated with an interaction graph $\Gamma$ with input vertices $I$, output vertices $J$ and measuring vertices $K$.

\smallskip
\noindent
{\bf Step $\>10\>$:}
A local von Neumann measurement is applied to the input vertices $I$ in $x$-basis and to a subset $K_1\subset K$ of measuring vertices in the product basis $\beta_1$ of $\H{K_1}$:
\begin{equation}\nonumber
P_{[\xi\otimes\beta_1,I K_1]}:\6C(I K_1)\otimes\6A(J K^1)\to \6A(I J K) 
\end{equation}
where $K^1=K\setminus K_1$ is the complement of $K_1$ in $K$. This measurement produces a random outcome $p^{I K_1}$.

\smallskip
\noindent
{\bf Step $\>11\>$:}
A conditional local unitary operation, subject to the outcomes of the previous measurement, is performed:
\begin{equation}\nonumber
C_1:\6A(J K^1)\to\6C(I K_1)\otimes\6A(J K^1) \; .
\end{equation}
It is implemented by the family of local unitary operators $\{\1s_{[C_1|p^{I K_1}]}|p^{IK_1}\in\7F^{IK_1}\}$ which are localized in $J K^1$, i.e. they belong to $\6A(JK^1)$.

\smallskip
\noindent
{\bf Step $\>t0\>$:}
In the $t$th measurement round a local von Neumann measurement operation with respect to a product basis $\beta_t$ of $\H{K_t}$ is applied at the measuring vertices $K_t\subset K^{t-1}$:
\begin{equation}\nonumber
P_{[\beta_t,K_t]}:\6C(K_t)\otimes\6A(J K^t)\to\6A(J K^{t-1}) \; .
\end{equation}
Here $K^t=K^{t-1}\setminus K_t$ is the complement of $K_t$ in $K^{t-1}$.

\smallskip
\noindent
{\bf Step $\>t1\>$:}
A local unitary operation, conditioned on the outcomes of the previous measurement, is performed:
\begin{equation}\nonumber
C_t:\6A(J K^t)\to\6C(K_t)\otimes\6A(J K^t) \; .
\end{equation}

\medskip

The wanted algorithm, is realized by proceeding the scheme inductively, until all qudits, placed at measurement vertices, are removed. Thus the channel $T$ is given by   
\begin{equation}
T=E_{I}\circ\alpha_\Gamma\circ F_0\circ\cdots\circ F_n \; ,
\end{equation}
where $F_t:=P_{[\beta_t,K_t]}\circ C_{t}$ is the composition of two elementary operations, a measurement and a corresponding conditional local unitary operation. 

The composition $\>00\>01\>$ of the initial steps $\>00\>$ and $\>01\>$ is interpreted as the ``reading in" of quantum information into the cluster. This information is ``spread out" through all qudits via the entanglement that is created by the dynamics. In fact, if the graph $\Gamma$ is connected, then the received states is ``highly entangled" in the sense of the {\em persistence of entanglement} \cite{BrieRau01b}. This entanglement can be used to run the quantum algorithm by an sequential application of local measurement operation. 

The composition of the steps $\>00\>01\>10\>$ (also symbolized by $\>00\to 10\>$) yields a channel which has an implementation by the Kraus operators
\begin{equation}\label{kraus-1}
\bigl\{ \ \Phi^*_{[p^{I K_1}]}u_1{u}(\Gamma)\Phi_{[{J K}]} \ \bigm|p^{IK_1}\in\7F^{IK_1} \ \bigr\} \; ,
\end{equation}
where $u_1$ is a local unitary operator that transforms the $x$-basis into the basis $\beta_1$. This channel is not a pure quantum channel since it produces classical measurement results $p^{I K_1}$ which are completely random. In order to ``control the procedure" of information processing, we have to require additional constraints on the conditional local unitary operation $C_1$ which is applied by step $\>11\>$. Controlling a procedure means here that we have no loss of information during the process. This requires that the channel $T$ is pure. By construction, the channel $T_{\>00 \to 11\>}$ which is given by the composition of the steps $\>00\to 11\>$ is implemented by the Kraus operators 
\begin{equation}\label{kraus-2}
\bigl\{ \ \1s_{[C_1|p^{I K_1}]}\Phi^*_{[p^{I K_1}]}u_1{u}(\Gamma)\Phi_{[{J K}]} \ \bigm | p^{IK_1}\in\7F^{IK_1}\ \bigr\} \; .
\end{equation} 
Purity implies that all elements in this set are multiples of one isometry. This is the case if the action of $C_1$ ``compensates the randomness" the effect of the measurement operation $P_{[\xi\otimes\beta_1,I K_1]}$. Whenever the measurement result $p^{I K_1}$ is received, the conditional unitary operations rotates the system back into the state which corresponds to the standard measurement outcome $p^{I K_1}=0^{I K_1}$. In other words, the implementation of $T_{\>00\to 11\>}$ is ``gauged" with respect to standard measurement results. Expressed in terms of the Kraus operators (\ref{kraus-1}) and (\ref{kraus-2}) this just means that the family of local unitary operators $\{\1s_{[C_1|p^{I K_1}]}|p^{IK_1}\in\7F^{IK_1}\}$ has to fulfill the identity
\begin{equation}\label{compensate}
\1s_{[C_1|p^{I K_1}]}\Phi^*_{[p^{I K_1}]}u_1{u}(\Gamma)\Phi_{[{J K}]}
=\Phi^*_{[{I K_1}]}u_1{u}(\Gamma)\Phi_{[{J K}]}
\end{equation}
for each measurement outcome $p^{I K_1}$. As a consequence the channel $T_{\>00\to 11\>}$ can be represented by the single isometric Kraus operator 
\begin{equation}\label{kraus-3}
\sqrt{d}^{|IK_1|}\Phi^*_{[{I K_1}]}u_1{u}(\Gamma)\Phi_{[{J K}]} \; .
\end{equation}
In order to realize the algorithm $T$, we are faced with the following two main tasks:

\begin{description}
\item[Task 1:]
For each measurement round $t$ we have to compute the local unitary operators $\1s_{[C_t|\cdot]}$ that compensates the randomness of the corresponding measurement.
\item[Task 2:]
The Kraus operators (\ref{kraus-3}), which are related to standard measurement results, have to be computed.
\end{description}

\paragraph*{\it Remark:}
The composed channel $F_1:=P_{[u_1\xi,I K_1]}\circ C_{1}$ can be interpreted as adjusting the basis for the next measurement round $P_{[\beta_2,K_2]}$ subject to the received measurement result of the previous round. More precisely, depending on the measurement result $p^{I K_1}$ of the first measurement round, the following local von Neumann measurement operation is performed in the adjusted product basis $\1s_{[C_1|p^{I K_1}]}\beta_2$.

If we view for the $t$th measurement round $\6A(I K_t)$ as Alice's observable algebra and $\6A(J K^t)$ as Bob's, then the channel $T_{\>00 \to t1\>}$ can be viewed as a teleportation scheme which propagates information through the cluster. After all qudits at $K$ has been measured, the information has been processed via ``teleporting" it from the input qudits into the output qudits. The entanglement within the cluster has been used up after the computation has been performed. This is the reason why the computational model, described above, is called a ``one-way quantum computer".

\section{Graph algorithms}
\label{graph-algo}
In the previous section the general scheme for implementing an algorithm via one-way quantum computing is discussed. The basic initial procedure is the initial preparation followed by an elementary step of the dynamics $\>00\>01\>$ which creates the basic entanglement resource. In principle the measurement operations, that follow, can be performed in any basis. The present section restricts to a particular class of bases that arise from stabilizer codes. They can be represented by graphs and the corresponding von Neumann measurements are called {\em graph measurements} as it is described in detail in the first part of this section.  

The algorithms under consideration make use of one local graph measurement round as it is described by Step~$\>10\>$. The first task is to {\em find an explicit construction for a conditional local unitary operation} that compensates the randomness of the graph measurement. It turns out that graph measurements can be converted into measurements in $x$-basis by an appropriate change of the underlying interaction graph. This enables for a particular class of graphs, which we call ``basic", to solve the task. In view of Step $\>11\>$, a family of Weyl operators, implementing the desired conditional local unitary operation, can be constructed explicitly. As a direct consequence, each encoding procedure of a graph code (and therefore each stabilizer code) can be implemented by a one-way quantum computer \cite{SchlWer00,Schl02}. 
 
Moreover, the second task can be solved, which consists in computing the Kraus operators for standard measurement outcomes. We show that standard measurement operations in $x$-basis can be viewed as operations on graphs. This leads not only to the fact that graph algorithms can be represented by one graph, which only has input and output vertices, we also obtain an explicit construction scheme that computs the final interaction graph from the underlying data, i.e. the given interaction and measuring graph.   

\subsection{Local graph measurements}
\label{local-graph-measurements}
A particular class of graph measurement operations ar the so called {\em graph measurements}. They are related to bases which emerge from stabilizer codes via Corollary \ref{cor-ew} in the following manner: We consider an admissible graph $\Lambda$ on the disjoint union $M K L$ of output vertices $M$, auxiliary vertices $K$ and syndrome vertices $L$. Due to Corollary \ref{cor-ew}, the set of vectors $\{\Psi_{[\Lambda|q^L]} |q^L\in\7F^L \}$ is an orthonormal basis of the Hilbert space $\H{M}$. The von Neumann measurement operation $P_{\Lambda}$, operating on the system $\6A(M)$, corresponds to this basis. It is the completely positive map which assigns to a classical observable $f\in\6C(L)$ the operator
\begin{equation}
P_{\Lambda}(f):=\sum_{q^L\in\7F_d^L} f(q^L) \ P_{[\Lambda|q^L]}
\end{equation}
in $\6A(M)$, where $P_{[\Lambda|q^L]}$ is the projection onto $\Psi_{[\Lambda|q^L]}$.  We also get a Kraus representation of $P_{\Lambda}$ by the co-isometries $\{\1v_{[\Lambda|q^L]}^*|q^L\in\7F^L \}$, where the operator $\1v_{[\Lambda|q^L]}^*$ is related to the measurement outcome $q^L$. Note that for an observable $A\in\6A(I)$ of a system, that is not affected by the measurement operation, we have $\1v_{[\Lambda|q^L]}A\1v_{[\Lambda|q^L]}^*=A\otimes P_{[\Lambda|q^L]}$.

Since the sets $M$ and $L$ contain the same number of elements, the set $M K L$ admits a partition into mutually disjoint sets
\begin{equation}
M K L=\bigcup_{m\in M} \{m\} \cup k(m) \cup \{l(m)\}
\end{equation}
where $m\mapsto l(m)$ is a bijective map which identifies the output vertices with the syndrome vertices and $k(m)$ is a set of auxiliary vertices. Suppose now the weighted graph $\Lambda$ is a sum of connected components $\Lambda=\sum_{M\in m}\Lambda_m$, where each $\Lambda_m$ is the admissible sub-graph on the vertices $\{m\}\cup k(m)\cup\{l(m)\}$. Then the vector $\Psi_{[\Lambda|q^L]}$ decomposes into a tensor product
\begin{equation}
\Psi_{[\Lambda|q^L]}=\bigotimes_{m\in M}\Psi_{[\Lambda_m|q^{l(m)}]} \;
\end{equation}
and $P_{\Lambda}$ is a local von Neumann measurement operation. The graph $\Lambda_m$ determines the measurement basis at the position $m$. 

\begin{figure}[h]
\vspace*{13pt}
\centerline{\epsfig{file=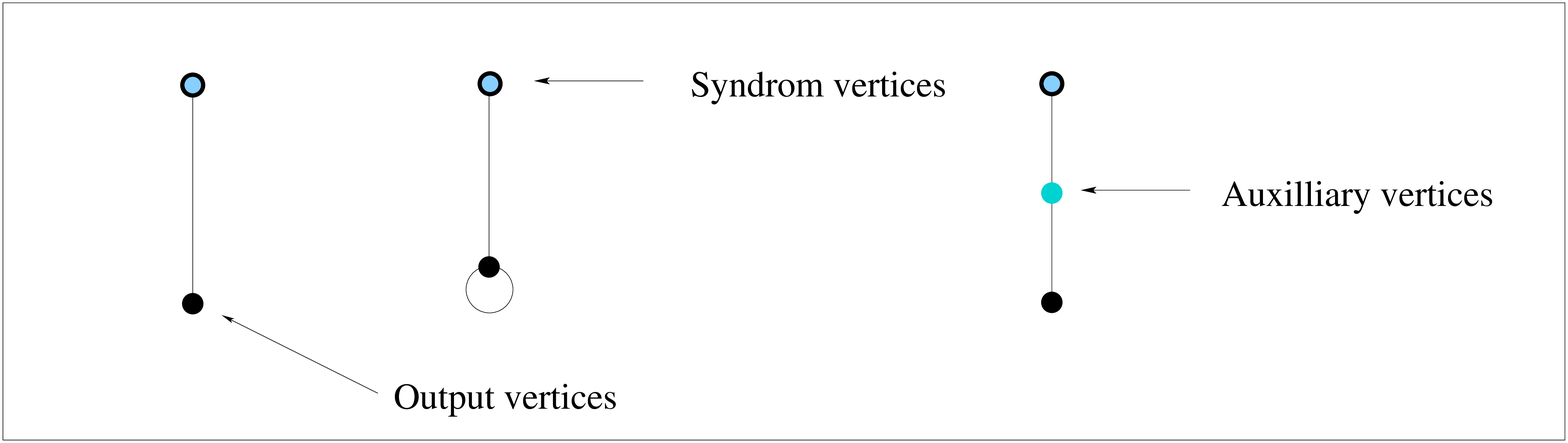, width=13cm}} 
\vspace*{13pt}
\fcaption{From left to right: The measurement graphs for local von Neumann measurements in $x$-, $y$-type- and $z$-basis.}
\label{graph}
\end{figure}

For analyzing graph measurements, it is sufficient to study graphs on two-elementary set $\{m,l\}$ (one output and one syndrome vertex) or three-elementary sets $\{m,k,l\}$. For the case that there are no auxiliary vertices, the simplest example for a measurement graph is given by the adjacency matrix
\begin{equation}
X=\left(\begin{array}{cc}
0&1_l^m\\
1_m^l&0
\end{array}\right)
\end{equation}
which represents the graph consisting of one single line that connects the vertex $j$ with $l$ (see Fig.~\ref{graph}). Here $1^n_m$ is the $1\times 1$-matrix block which corresponds to one edge that connects the vertices $n$ and $m$. Obviously, $X$ is admissible. The orthonormal basis, that consists of the vectors  $\Psi_{[X|q^l]}={\xi}_{[1^m_lq^l]}$, $q^l\in\7F$, is just the $x$-basis since
\begin{equation}
\Psi_{[X|q^l]}(q^m)=\chi(1^m_lq^l|q^m)={\xi}_{[1^m_lq^l]}(q^m) \; .
\end{equation}
The corresponding stabilizer algebra $\6A(m|X)$ is generated by shift operators $\1w(X^m_mq^m|q^m)=\1x(q^m)$.

A further example, which is almost as simple is given by the adjacency matrix
\begin{equation}
Y=\left(\begin{array}{cc}
n&1_l^m\\
1_m^l&0
\end{array}\right)
\end{equation}
of the admissible ``tadpole" graph consisting of one line that connects the vertex $k$ with $l$ and $n$ self-links at the vertex $m$ (see also Fig.~\ref{graph}). In this case, the orthonormal basis, which is given by the vectors $\Psi_{[Y|q^l]}$, $q^l\in\7F$, is a $y$-type
basis:
\begin{equation}
\Psi_{[Y|q^l]}(q^m)=\tau(n|q^m)\chi(1^m_lq^l|q^m)\; .
\end{equation}
The corresponding stabilizer algebra $\6A(m|Y)$ is generated by Weyl operators $\1w(nq^m|q^m)$, i.e. the vectors $\Psi_{[Y|q^l]}$, $q^l\in\7F$, form a basis of eigenvectors for the Weyl operators $\1w(n q^m|q^m)$. Note, that for $n=0$ we obtain the $x$-basis.

The following example makes use of an auxiliary vertex. The adjacency matrix
\begin{equation}
Z=\left(\begin{array}{ccc}
0&1_k^m&0\\
1_m^k&0&1_l^k\\
0&1^l_k&0
\end{array}\right)
\end{equation}
corresponds to the measurement graph which connects $m$ with $k$ and $k$ with $l$ (see Fig.~\ref{graph}). Clearly, $Z$ is admissible. Now, the orthonormal basis, consisting of the vectors   $\Psi_{[Z|q^l]}={\zeta}_{[1^m_lq^l]}$, $q^l\in\7F$, is the $z$-basis since
\begin{equation}
\Psi_{[Z|q^l]}(q^m)=\sqrt{d}\int\8dq^k \ \chi(q^k|1^k_mq^m+1^k_lq^l)={\zeta}_{[1^m_lq^l]}(q^m) \; .
\end{equation}
Indeed, the kernel of the block
\begin{equation}
Z^{k}_{\{m,k\}}=\left(\begin{array}{ccc}
1_m^k&0
\end{array}\right)
\end{equation}
consists of those vectors $q^{\{m,k\}}$, for which the component $q^m=0$ vanishes. Thus the corresponding stabilizer algebra $\6A(m,k|Z)$ is generated by shift operators
$\1w(1_k^m q^k|0^m)=\1z(1_k^m q^k)$. As a consequence, all measurement procedures in $x$-, $y$-type and $z$-bases are covered by considering graph measurements. 

Now we consider an applicaion of a graph measurement for the first round (Step $\>10\>$). The corresponding algorithm is called a {\em graph algorithm}. It is constructed from a {\em interaction graph} $\Gamma$, with input vertices $I$, output vertices $J$ and measuring vertices $M$, and from a {\em local measuring graph} $\Lambda$ which lives on $M$, the auxiliary vertices $K$ and the syndrome vertices $L$. The channel $T_{[\Gamma|\Lambda]}$, which describes the algorithm, is defined by the following sequence of elementary operations (See Fig.~\ref{alg-graph}): 
\begin{figure}[h]
\vspace*{13pt}
\centerline{\epsfig{file=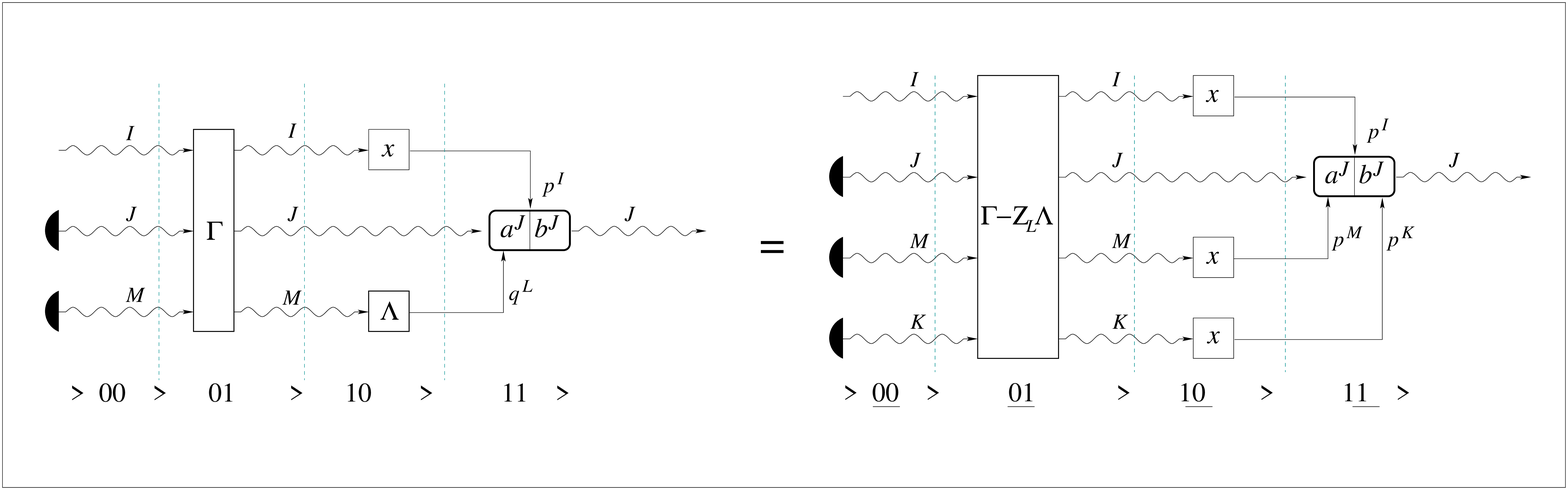, width=13cm}} 
\vspace*{13pt}
\fcaption{The graph algorithm for an interaction graph $\Gamma$ and a measurement graph $\Lambda$ is represented on the left hand side. The sequence of the single elementary steps $\>00\to 11\>$ is indecated below the figure. Both four step procedures $\>00 \to 11\>$ and $\>\underline{00}\to\underline{11}\>$ yield the same operation. It realizes the graph code associated with the graph $\Gamma-\8Z_L\Lambda$.}
\label{alg-graph}
\end{figure}

\smallskip
\noindent
{\bf Step $\>00\>$:} 
The quantum information, we wish to process is encoded via the operation $E_I$, i.e. each qudit at the positions $I M$ is prepared in the standard state. 

\smallskip
\noindent
{\bf Step $\>01\>$:} 
An elementary step of the dynamics $\alpha_\Gamma$ is applied, where the graph $\Gamma$ lives on the union of input vertices $I$, measuring vertices $M$, and output vertices $J$.

\smallskip
\noindent
{\bf Step $\>10\>$:}
The input qudits are measured in $x$-basis $P_{[\xi,I]}$. At the measurement vertices $M$, a graph measurement $P_{\Lambda}$ is performed for a measuring graph $\Lambda$, with measuring vertices  $M$, auxiliary vertices $K$ and syndrome vertices $L$. This produces measurement outcomes $p^I,q^{L}$. 

\smallskip
\noindent
{\bf Step $\>11\>$:} 
A conditional local unitary operation $C_{[\Gamma,\Lambda]}$ is applied that compensates the randomness of measurement $P_{[\xi,I]}\otimes P_\Lambda$. 

\medskip

Let us first concentrate on the channel $T_{\>00\>01\>10\>}$ which is given by composing the steps $\>00\>01\>10\>$ (See Fig.~\ref{alg-graph} for illustration). It has a representation by the set 
\begin{equation}\label{kraus-6}
\bigl\{ \ \sqrt{d}^{-|I|}\1v_{[\Lambda|q^L]}^*{\1u}_{[\Gamma|p^I]} \ \bigm| q^L\in\7F^L\bigr\} 
\end{equation} 
of Kraus operators. For convenience, we have introduced here the {\em encoding operator}
\begin{equation}
{\1u}_{[\Gamma|p^N]}:=\sqrt{d}^{|N|}\Phi_{[p^N]}^*{u}(\Gamma)\Phi_{[{J M}]}\; ,
\end{equation}
which is associated with the graph $\Gamma$ and the measurement result $p^N$, where $N$ is a subset in $I M$ that consists of input and measuring vertices. As we shall see later, the encoding operator associated with $\Gamma$ for the standard measurement outcome ${\1u}_{[\Gamma|N]}:={\1u}_{[\Gamma|0^N]}$ plays a particular role: The encoding operator, associated with the graph $\Gamma$ and the standard measurement outcome, is the Kraus operator that implements the quantum code associated with $\Gamma$ in the sense of \cite{SchlWer00,Schl02}. 

So far everything can be expressed explicitly in terms of the given interaction graph $\Gamma$ and measuring graph $\Lambda$. The problem, which occurs next, is to given an explicit construction for the operation $C_{[\Gamma,\Lambda]}$. At least we have to show the existence of local unitary operators $\1s_{[\Gamma,\Lambda|p^I,q^L]}$ which compensate the measurement outcomes $p^I,q^L$:
\begin{equation}\label{graph-compensate}
\1v_{[\Lambda|0^L]}^*{\1u}_{[\Gamma|I]}
=\1s_{[\Gamma,\Lambda|p^I,q^L]}\1v_{[\Lambda|q^L]}^*{\1u}_{[\Gamma|p^I]} \; .
\end{equation} 
If we can solve the Equation (\ref{graph-compensate}) for each measurement outcome $p^I,q^L$, then we know that the channel $T_{[\Gamma|\Lambda]}$  has a representation in terms of the single isometric Kraus operator 
\begin{equation}\label{kraus-7}
\sqrt{d}^{|L|}\1v_{[\Lambda|0^L]}^*{\1u}_{[\Gamma|I]}\; .
\end{equation}

In order to construct explicit solutions of the equations (\ref{graph-compensate}), we make use of the fact that every graph measurement can be realized by just performing measurements in $x$-basis. More precisely:
\begin{thm}\label{x-measure}
For each measurement outcome $(p^I,q^L)$ the identity
\begin{equation}
\1v_{[\Lambda|q^L]}^*{\1u}_{[\Gamma|p^I]}
=
\sqrt{d}^{-|L|}{\1u}_{[\Gamma-\8Z_L\Lambda|p^I-\Lambda^{M K}_Lq^L]}
\end{equation}
holds, where $\8Z_L\Lambda=\Lambda^{M K}_{M K}$ is the sub-graph which is obtained via removing the vertices $L$.
\end{thm}
\proof{Lemma~\ref{tech-lem-0} provides a useful relation between the isometry $\1v_{[\Lambda|q^L]}$ with the encoding operator ${\1u}_{[\Gamma|p^I]}$. Namely, 
\begin{equation}
\1v_{[\Lambda|q^L]}^*{\1u}_{[\Gamma|p^I]}=\Pi_{[q^L]}^*\1u_{[\Gamma-\Lambda|0^{M K},p^I]}
\end{equation}
holds, where $\Gamma-\Lambda$ is a graph with input vertices $I$, measuring vertices $K M$   
and output vertices $J L$. Due to Lemma~\ref{tech-lem-5} it follows that encoding operator, associated with the graph $\Gamma-\Lambda$ and the measurement outcome $(0^{M K},p^I)$ fulfills the identity 
\begin{equation}
\1u_{[\Gamma-\Lambda|0^{M K},p^I]}=\tau(\Lambda|q^L)\1w(\Lambda^L_Lq^L|q^L)
\1u_{[\Gamma-\Lambda|p^I-\Lambda^{K M}_Lq^L]} \; .
\end{equation}
We apply the co-isometry $\Pi_{[q^L]}^*=\Pi_{[L]}^*\1x(-q^L)$ on both sides of this equation and we exploit the fact that there are no edges of $\Lambda$ which connect syndrome vertices $L$. This yields the relation
\begin{equation}
\1v_{[\Lambda|q^L]}^*{\1u}_{[\Gamma|p^I]}
=\Pi_{[L]}^*\1u_{[\Gamma-\Lambda|p^I-\Lambda^{K M}_Lq^L]} \; .
\end{equation}
One observes that for each function $\psi\in\H{L J}$ the identity $(\Pi^*_{[L]}\psi)(q^J)=\sqrt{d}^{-|L|}\psi(0^L,q^J)$ holds which finally implies that
\begin{equation}
\Pi_{[L]}^*\1u_{[\Gamma-\Lambda|p^I-\Lambda^{K M}_Lq^L]}=\sqrt{d}^{-|L|} \ \1u_{[\Gamma-\8Z_L\Lambda|p^I-\Lambda^{K M}_Lq^L]}
\end{equation}
is true. This concludes the proof of the theorem}

One immediate consequence of the theorem is that the channel $T_{[\Gamma|\Lambda]}$ can be decomposed by an alternative sequence of elementary operations which only include local measurements in $x$-basis (See Fig.~\ref{alg-graph}): 

\smallskip
\noindent
{\bf Step $\>\underline{00}\>$:} 
The quantum information, we wish to process is encoded via the operation $E_I$, i.e. each qudit at the positions $J K M$ is prepared in the standard state.

\smallskip
\noindent
{\bf Step $\>\underline{01}\>$:} 
An elementary step of the dynamics $\alpha_{[\Gamma-\8Z_L\Lambda]}$ is applied, associated with the graph $\Gamma-\8Z_L\Lambda$ on the union of input vertices $I$, measuring vertices $M K$, and output vertices $J$. 

\smallskip
\noindent
{\bf Step $\>\underline{10}\>$:}
A measurement in $x$-basis $P_{[\xi,I M K]}$ is performed at the input vertices $I$ and the measuring vertices $M K$, producing a measurement outcome $p^{I M K}$. 

\smallskip
\noindent
{\bf Step $\>\underline{11}\>$:} 
The randomness of the previous measurement is compensated by a conditional local unitary operation $C_{[\Gamma-\8Z_L\Lambda]}$.  

\medskip

To verify the equivalence of both procedures, we make use of the observation, that the composition of the steps $\>\underline{00}\>\underline{01}\>\underline{10}\>$ has a representation by the Kraus operators 
\begin{equation}
\bigl\{ \ \sqrt{d}^{-|IMK|}\1u_{[\Gamma-\8Z_L\Lambda|p^{I M K}]} \ \bigm| p^{IMK}\in\7F^{IMK} \bigr\}
\end{equation}
each corresponding to the measurement outcome $p^{I M K}\in\7F_d^{I M K}$. If we assert here that the randomness of the measurement outcomes can be compensated, then for  the channel $T_{[\Gamma-\8Z_L\Lambda]}$, which is given by composing the steps $\>\underline{00}\to\underline{11}\>$, is represented by the single isometric Kraus operator $\1u_{[\Gamma-\8Z_L\Lambda|{I M K}]}$.
By Theorem \ref{x-measure}, this isometry coincides with the Kraus operator (\ref{kraus-7}) and therefore both procedures yields the same result (Fig.~\ref{alg-graph}, right hand side): $T_{[\Gamma-\8Z_L\Lambda]}=T_{[\Gamma|\Lambda]}$. Therefore it is just one graph $\Gamma-\8Z_L\Lambda$ needed for determining a graph algorithm.

\subsection{Compensating the randomness}
\label{subsec-comp}
As discussed in the previous subsection, all algorithms which are based on graph measurements can be converted into algorithms which are only faced with measurements in $x$-basis. Without loss of generality, each graph algorithm can be represented by one graph $\Gamma$ with input vertices $I$, measuring vertices $K$ and output vertices $J$. The corresponding channel $T_\Gamma$ is then constructed by composing the four elementary operations $\>\underline{00}\to\underline{11}\>$ given in the previous section, where the graph $\Gamma-\8Z_L\Lambda$ is just replaced by $\Gamma$.

Since we are concerned with $x$-basis measurements we expect, following the analysis of \cite{BrieRau01c}, that the local unitary ``byproduct" operators, which implement $C_\Gamma$, are Weyl operators. In view of this we introduce the following notion: A linear map $\Theta$ which assigns  to each measurement outcome $p^{I K}$ a phase space vector (translation) $\Theta p^{I K}\in\Xi^J$ is called {\em compensating} for the measurement operation in $x$-basis at $I K$ if the equation
\begin{equation}\label{compensator}
{\1u}_{[\Gamma|{I K}]}=\lambda(p^{I K})\1w(\Theta p^{I K})^*{\1u}_{[\Gamma|p^{I K}]}
\end{equation}
holds for all measurement outcomes $p^{I K}$, where $\lambda(p^{I K})$ is some phase depending on $p^{I K}$. 

If we assume that there exists a compensating linear map $\Theta$, then the channel $C_\Gamma$ is implemented by the Weyl operators 
\begin{equation}
\1s_{[\Gamma|p^{I K}]}=\1w(\Theta p^{I K}) \; .
\end{equation}
But what can one say about the existence of compensating linear maps yielding appropriate conditional local unitary operations? As already discussed above, it follows from the fact that the linear map $\Theta$ is compensating, that $T_\Gamma$ is implemented by a single isometric Kraus operator, namely the encoding operator $\1u_{[\Gamma|I K]}$. Thus the following proposition is a necessary condition:

\begin{prop}
If there exists a compensating linear map for the measurement in $x$-basis at $I K$, then the encoding operator ${\1u}_{[\Gamma|{I K}]}$ is isometric.  
\end{prop}

Sufficient conditions for the existence as well as an explicit construction of compensating linear maps, can be formulated in terms of the interaction graph: An graph on the union of input vertices $I$, measurement vertices $K$, and output vertices $J$, is called a {\em basic graph} if the block matrix $\Gamma^{K J}_{I K}$ is injective. The set of all basic graphs $\Gamma$ is denoted by $\8{BG}(I,J;K)$.

We point out here that the condition for a graph $\Gamma$ to be basic implies that the encoding operator ${\1u}_{[\Gamma|{I K}]}$ is an isometry and vice versa. We show that the following is true:

\begin{thm}\label{thm-byprod}
Let $\Gamma$ be weighted graph on the union of input $I$, output $J$, and measuring vertices $K$. There exists a compensating linear map $\Theta$ for the measurement in $x$-basis at $I K$ if and only if $\Gamma\in\8{BG}(I,J;K)$ is a basic graph and $\Theta$ is given by
\begin{equation}\label{eq-simpl-comp}
\Theta=(\Gamma^{J}_{K J}\bar\Gamma^{K J}_{I K}|-\bar\Gamma^J_{I K})
\end{equation}
where  $\bar\Gamma^{K J}_{I K}$ is a right inverse of the block matrix $\Gamma_{K J}^{I K}$, that is 
$\Gamma_{K J}^{I K}\bar\Gamma^{K J}_{I K}=1^{I K}_{I K}$.
\end{thm}
\proof{In the appendix, we prove Lemma \ref{tech-lem-4-a} which states that for a graph $\Gamma$ with input vertices $I$, output vertices $J$ and measuring vetices $K$, the identity  
\begin{equation}\label{comp-app-1}
{\1u}_{[\Gamma|{I K}]}
=\tau(\Gamma|q^{K J})\1w(-\Gamma^{J}_{K J}q^{K J}|q^{J})^*{\1u}_{[\Gamma|p^{I K}]}
\; .
\end{equation}
holds if the equation
\begin{equation}\label{eq-compensate-1}
\Gamma^{I K}_{K J}q^{K J}+p^{I K}=0
\end{equation}
is valid.  If we assume that $\Gamma$ is basic, then for each measurement outcome $p^{I K}$ the equation (\ref{eq-compensate-1}) has a solution $q^{K J}$ which is given by 
\begin{equation}
q^{K J}=-\bar\Gamma_{I K}^{K J}p^{I K}
\end{equation}
where $\bar\Gamma_{I K}^{K J}$ is a right inverse for $\Gamma^{I K}_{K J}$ which exists since the block matrix $\Gamma^{I K}_{K J}$ is surjective (note that $\Gamma_{I K}^{K J}$ is injective). By inserting this solution into (\ref{comp-app-1}) we obtain 
\begin{equation}\label{comp-app-1}
{\1u}_{[\Gamma|{I K}]}
=\tau(\Gamma|-\bar\Gamma_{I K}^{K J}p^{I K})\1w(\Gamma^{J}_{K J}\bar\Gamma_{I K}^{K J}p^{I K}|-\bar\Gamma_{I K}^{J}p^{I K})^*{\1u}_{[\Gamma|p^{I K}]}
\; .
\end{equation}
Thus we conclude that $\Theta=(\Gamma^{J}_{K J}\bar\Gamma^{K J}_{I K}|-\bar\Gamma^J_{I K})$ is indeed compensating. Vice versa, if we assume that there exists a compensating map, then ${\1u}_{[\Gamma|I K]}$ is an isometry. By Lemma~\ref{tech-lem-00}, which is also proven in the appendix, it follows that $\Gamma$ is basic}

As already mentioned, for a basic graph $\Gamma\in\8{BG}(I,J;K)$ the isometry $\1u_{[\Gamma|I K]}$ is the encoding isometry of the graph code associated with $\Gamma$ \cite{SchlWer00,Schl02}. Now we conclude from Theorem~\ref{thm-byprod}:

\begin{cor}
Each stabilizer code can be implemented on a one-way quantum computer via an algorithm which is composed of the elementary steps $\>\underline{00}\to\underline{11}\>$.
\end{cor}

\subsection{Standard measurements and operations on graphs}
As long as we are concerned with basic graphs $\Gamma\in\8{BG}(I,J;K)$ the statement of Theorem \ref{thm-byprod} tells us that the randomness of a measurement in $x$-basis at $I K$ can be compensated via a conditional local unitary operation. In other words, we can solve the ``first task" that goes along with one-way quantum computing. Thus for the implementation of the algorithm $T_{\Gamma}$ it remains to tackle the second task which consists in computing the standard measurement operations. For this purpose, we simplify the channel $T_{\Gamma}$ by removing the measurement vertices $K$ such that we are left with input and output vertices only. In fact, it turns out that the standard measurement operations at $K$ can be interpreted as operations on graphs. 
 
For a basic graph $\Gamma\in\8{BG}(I,J;K)$ we call a subset $N\subset J K$ of measuring and output vertices {\em pre-removable} with respect to $\Gamma$ if the block matrix $\Gamma_{N}^{N}$ is invertible, where $\bar\Gamma_{N}^{N}$ denotes the inverse. We write $\8{PR}(I,J,K|\Gamma)$ for the collection of all pre-removable sets. If a pre-removable set $N$ contains no output vertices, we call it {\em removable}.

Let $M$ be a finite set which contains the same number of elements as $J\cap N$. Let $\nu\mathpunct:M\to J\cap N$ be a bijective map. Then we denote by $\widehat{\nu}$ the connecting graph on $(J\cap N) M$ which connects the vertex $m\in M$ with $\nu(m)\in J\cap N$ by one line. We obtain a new graph $\Gamma+\widehat{\nu}$ on the vertices $I J K M$ which is, in particular, a basic graph with input vertices $I$, output vertices $(J\setminus N) M$ and measuring vertices $N K$. For this graph the block matrix
\begin{equation}
(\Gamma+\widehat{\nu})^N_N=\Gamma^N_N+\widehat{\nu}^N_N=\Gamma^N_N
\end{equation}
is invertible. $N$ is removable for $\Gamma+\widehat{\nu}$, and we can build the {\em Schur complement}
\begin{equation}\label{schur-compl}
\8X_N(\Gamma+\widehat\nu)
=\Gamma^{\bar N}_{\bar N}
-(\Gamma^{\bar N}_{N}+\widehat{\nu}^{M}_{J\cap N})\bar\Gamma^{N}_{N}
(\Gamma_{\bar N}^{N}+\widehat{\nu}_{M}^{J\cap N})
\end{equation}
which is a graph with vertices $\bar N M$, where $\bar N$ is the complement of $N$ in $I J K$. 

An important observation is here, that, if the underlying graph $\Gamma$ is basic, then the graph $\8X_N(\Gamma+\widehat\nu)$ is also basic. Namely, for any basic graph $\Lambda$ with input vertices $I$, output vertices $J$ and measuring vertices $K$ the following holds: Suppose that $N\subset K$ is removable. The Schur complement $\8X_N\Lambda$ is a graph with input vertices $I$, output vertices $J$ and measuring vertices $M=K\setminus N$. In order to show that $\8X_N\Lambda$ is a basic graph we study the kernel of the block matrix $(\8X_N\Lambda)^{J M}_{I M}$ which is nothing else but analysing the system of equations
\begin{equation}\label{kernel-sc}
(\8X_N\Lambda)^{J M}_{I M}q^{I M}= \Lambda^{J M}_{I M}q^{I M}
-\Lambda^{J M}_{N} \bar\Lambda^{N}_{N} \Lambda_{I M}^{N}q^{I M}=0
\end{equation} 
which can equivalently be written as
\begin{equation}
\Lambda^{J M}_{I K}(q^{I M}-\bar\Lambda^{N}_{N} \Lambda_{I M}^{N}q^{I M})=0 \; .
\end{equation} 
Since $\Lambda$ is basic graph, it follows: If $q_1^{I K}$ belongs to the kernel of $\Lambda^{N}_{I K}$, then $\Lambda^{J M}_{I K}q_1^{I K}=0$ implies that $q_1^{I K}=0$. Thus $q^{I M}=0$ follows from (\ref{kernel-sc}) if $q_1^{I K}=q^{I M}-\bar\Lambda^{N}_{N} \Lambda_{I M}^{N}q^{I M}$ is contained in the kernel of $\Lambda^N_{I K}$. Indeed we find that  
\begin{equation}
\Lambda^N_{I K}(q^{I M}-\bar\Lambda^{N}_{N} \Lambda_{I M}^{N}q^{I M})
=\Lambda^N_{I M}q^{I M}
-\Lambda^N_{N}\bar\Lambda^{N}_{N}\Lambda_{I M}^{N}q^{I M} 
=0
\end{equation}
holds. Thus we have shown that the Schur complement of a basic graph is again a basic graph. Now we can apply this fact to the graph $\Lambda=\Gamma+\widehat{\nu}$ which proves that $\8X_N(\Gamma+\widehat{\nu})$ is indeed basic.      

The prescription $\Gamma\mapsto\8X_N(\Gamma+\widehat{\nu})$ is an operation on graphs which leaves the number of input vertices as well as the number of output vertices fixed, because $|J\cap N|=|M|$ and thus $|J\setminus N|+|M|=|J|-|J N|+|M|=|J|$. On the other hand, the number of measuring vertices is reduced by $|K\cap N|$. We shall see, that standard measurement operations in $x$-basis, performed at a subset $N\cap K$ of measuring vertices $K$, can directly be viewed as applying the map $\Gamma\mapsto\8X_N(\Gamma+\widehat{\nu})$ to the underlying graph. 

\begin{thm}\label{thm-remove}
Let $\Gamma\in\8{BG}(I,J;K)$ be a basic graph, let $N\in\8{PR}(I,J,K|\Gamma)$ be a pre-removable set and let $\nu\mathpunct:J\cap N\to M$ be a bijective map. Then the identity
\begin{equation}\label{remove}
F_{[\widehat{\nu}^{M}_{J\cap N}]} \ {\1u}_{[\Gamma|{I K}]}=
\kappa \ {\1u}_{[\8X_N(\Gamma+\widehat{\nu}) |{I K\setminus N}]}
\end{equation}
holds for some phase $\kappa$, where $F_{[\widehat{\nu}^{M}_{J\cap N}]}$ is the local Fourier transform with respect to the connecting graph $\widehat{\nu}$ acting at the output vertices $J\cap N$.
\end{thm}
\proof{
We first apply Lemma~\ref{tech-lem-6}, given in the appendix, to the basic graph $\Gamma+\widehat\nu$ which has input vertices $I$, output vertices $(J\setminus N) M$ and measuring vertices $N K$. The set $N$ is removable for the graph $\Gamma+\widehat\nu$ and the operator  
\begin{equation}\label{multiple}
{\1u}_{[\Gamma+\widehat\nu|I N K]}^*{\1u}_{[\8X_N(\Gamma+\widehat\nu)|I K\setminus N]}\in\7C\11_I
\end{equation}
is a multiple of the identity in $\6A(I)$ since it can be shown that the operator (\ref{multiple}) commutes with all Weyl operators with help of Lemma~\ref{tech-lem-5}. Thus we conclude that there is a phase $\kappa$ such that    
\begin{equation}\label{remove-1}
{\1u}_{[\Gamma+\widehat\nu|I N K]}=\kappa \ {\1u}_{[\8X_N(\Gamma+\widehat\nu)|I K\setminus N]}
\end{equation}
holds. Note that $\kappa$ has to be a phase since both operators, ${\1u}_{[\Gamma+\widehat\nu|I N K]}$ and ${\1u}_{[\8X_N(\Gamma+\widehat\nu)|I K\setminus N]}$ are isometries. Finally we employ Lemma~\ref{tech-lem-7} which states that the identity 
\begin{equation}\label{fourier-1}
F_{[{\widehat\nu}_N^M]}{\1u}_{[\Gamma|I K]}
={\1u}_{[\Gamma+\widehat\nu|I K N]}
\end{equation}
holds. This means that the composition of the Fourier transform $F_{[{\widehat\nu}_N^M]}$ with the isometry ${\1u}_{[\Gamma|I K]}$ can be viewed as joining the vertices of the connecting graph $\widehat\nu$ to the vertices of $\Gamma$ and adding the edges of both graphs. Combining the equations (\ref{remove-1}) and (\ref{fourier-1}) yields the desired identity (\ref{remove})}

The identity (\ref{remove}) simplifies, if we just consider a removable set $N\in\8{PR}(I,J,K|\Gamma)$. It is not necessary to make use of a bijective map $\nu$ and the Fourier transform can be skipped. The Schur complement for $\Gamma$ is then given by:

\begin{cor}
Let $\Gamma\in\8{BG}(I,J;K)$ be a basic graph. If the set $N\subset K$ is removable, then the identity
\begin{equation}
{\1u}_{[\Gamma|{I K}]}=\kappa \ {\1u}_{[\8X_{N}\Gamma|{I K\setminus N}]}
\end{equation}
holds for some phase $\kappa$.
\end{cor}

Theorem~\ref{thm-remove} can be used to remove all measurement vertices in a systematic way by just considering one- and two-elementary sets. The following proposition is a further useful ingredient for this purpose.

\begin{prop}\label{cases}
Let $\Gamma\in\8{BG}(I,J,K)$ be a basic graph. Then for each measuring vertex $n\in K$, one of the
following statements is true:
\begin{description}
\item[\rm (i)]
The vertex $n$ has a self-link, $\Gamma(n,n)\not=0$ which implies that the set $\{n\}$ is removable.

\item[\rm (ii)]
The vertex $n$ has no self-link, $\Gamma(n,n)=0$ but it is connected to another measuring vertex $n'\in K$, i.e. $\Gamma(n,n')\not=0$. This implies that the set $N=\{n,n'\}$ is removable.

\item[\rm (iii)]
The vertex $n$ has no self-link, $\Gamma(n,n)=0$ but it is connected to an output vertex $j\in J$, i.e.
$\Gamma(n,j)\not=0$. This implies that the set $N=\{n,j\}$ is pre-removable.
\end{description}
\end{prop}
\proof{
We choose a measuring vertex $n\in K$. The statement corresponding to case (i) is obvious. Suppose now that $m$ is a measuring or an output vertex in $J K$ which is connected to $n$. This implies indeed that the set $N=\{n,m\}$ is removable since the determinant of the block matrix $\Gamma^N_N$ is non-vanishing $\det(\Gamma^N_N)=\Gamma(n,n)\Gamma(m,m)-\Gamma(n,m)^2=-\Gamma(n,m)^2\not=0$ and it follows that $\Gamma^N_N$ is invertible, i.e. the statements (ii) and (iii) hold. To complete the proof we have to exclude the case where the measuring vertex $n$ is connected to non of the measuring or an output vertices. In this case we find for a register configuration $q^{I K}=q^n\not=0$, which only has a non-zero entry at position $n$, that the identity  $\Gamma^{J K}_{I K}q^{I K}=\Gamma^{J K}_{n}q^{n}=0$ holds. This contradicts the assumption that $\Gamma$ is a basic graph, i.e. the kernel of $\Gamma^{J K}_{I K}$ is trivial}  

We can derive from the results of the previous discussion an inductive algorithm for removing the measuring vertices of a basic graph $\Gamma\in\8{BG}(I,J;K)$. Suppose after $s$ steps, we have removed the measuring vertices $M_s$ such that we are left with $K_s=K\setminus M_s$ remaining measuring vertices and a basic graph $\Gamma_s\in\8{BG}(I,J_s,K_s)$. The set of output vertices $J_s$
has the same cardinality as $J$ and consists of a part $J_s\cap J$ of initial output vertices and a part $J_s\setminus J$ of exchanged output vertices. Then we remove a further measuring vertex $n\in K_s$ conditioned by one of the possibilities, given by Proposition~\ref{cases}.

We also have to keep in mind that some operations on graphs has to be compensated by a product of local Fourier transformations $F_s\mathpunct:\H{J_s}\to\H{J}$ which operate trivially on the qudits with positions $J\cap J_s$.

\begin{description}
\item[\rm (i)]
If the vertex $n$ has a self-link, $\Gamma_s(n,n)\not=0$, it is removable and we build the basic graph
\begin{equation}\label{schur-comp-y}
\Gamma_{s+1}:=\8X_{n}\Gamma_s
\end{equation}
with input vertices $I$, output vertices $J_{s+1}=J_s$ and measuring vertices $K_{s+1}:=K_s\setminus n$. The Fourier transform  $F_{s+1}=F_s$ remains unchanged.

For the binary case, the Schur complement (\ref{schur-comp-y}) can be obtained by a useful graphical role: Take the subgraph by removing the vertex $n$. Add further edges to it by mutually connecting those vertices which are linked to $n$, including self links. Note that edges are added modulo $2$. Fig.~\ref{remove-intro-1} represents this procedure.

\item[\rm (ii)]
The vertex $n$ has no self-link, but it is connected to another measuring vertex $k\in K_s$, i.e. $\Gamma_s(n,k)\not=0$. We build the basic graph
\begin{equation}\label{schur-comp-x}
\Gamma_{s+1}:=\8X_{\{n,k\}}\Gamma_s
\end{equation}
with input vertices $I$, output vertices $J_{s+1}=J_s$ and measuring vertices $K_{s+1}:=K_s\setminus\{n,k\}$. The Fourier transform  $F_{s+1}=F_s$ remains unchanged.

Again we have a graphical role for obtaining (\ref{schur-comp-x}) for the binary case. By following the strategy to remove all measurement vertices with self-links first, we may assume without loss of generality that both vertices $n$ and $k$ have no self-links. Then we can proceed as follows: Build the subgraph by removing the vertices $n$ and $k$. Then add edges by connecting each vertex, which is linked to $n$, with every vertex that is linked to $k$ (See Fig.~\ref{remove-intro-2} for illustation).  

\item[\rm (iii)]
The vertex $n$ has no self-link, but it is connected to an output vertex $k\in J_s$. Let $m$ be an additional vertex. Then $1^m_l+1^l_m$ is the adjacency matrix of the graph that connects the vertex $l$ with $m$ by one line. We build the basic graph
\begin{equation}
\Gamma_{s+1}:=\8X_{\{n,k\}}(\Gamma_s+1^m_k+1^k_m)
\end{equation}
with input vertices $I$, output vertices $J_{s+1}=(J_s\setminus k) m$ and measuring vertices $K_{s+1}:=K_s\setminus n$. The new Fourier transform  is given by $F_{s+1}=F_sF[1^k_m]$.

Concerning binary systems, we may use here the same graphical role as described for case (ii). 
\end{description}

\paragraph*{\it Example I:}
\begin{figure}[h]
\vspace*{13pt}
\centerline{\epsfig{file=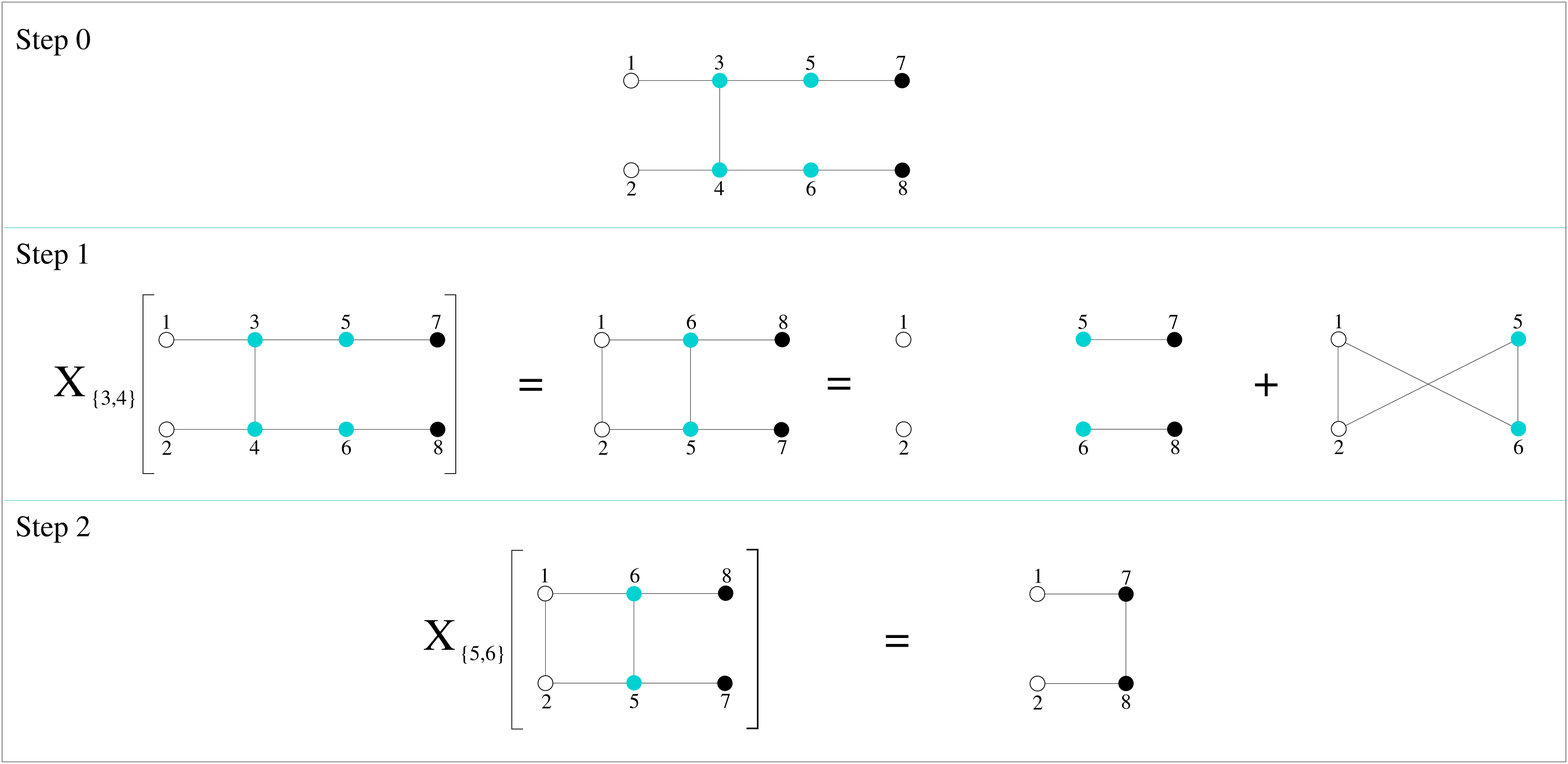, width=13cm}} 
\vspace*{13pt}
\fcaption{Step~0: The initial graph $\Gamma_0$. Step~1: The set measuring vertices $\{3,4\}$ is removable. According to (ii), we obtain a new graph $\Gamma_1$ with two connected measuring vertices $\{5,6\}$. Step~2: The set measuring vertices $\{5,6\}$ is removable. According to (ii), we finally obtain the graph $\Gamma_2$ which has no measuring vertices left.}
\label{step-2}
\end{figure}
Fig.~\ref{step-2} shows a simple example for a qubit system ($d=2$). The initial graph $\Gamma_0$ has input vertices $\{1,2\}$, measuring vertices $\{3,4,5,6\}$, and output vertices $\{7,8\}$. The adjacency matrix is given by
\begin{equation}
\Gamma_0=\left(\begin{array}{cc|cccc|cc}
 0 &  0 &  1 &  0 &  0 &  0 &  0 &  0 \\
0 &  0 &  0 &  1 &  0 &  0 &  0 &  0 \\
\hline
      1 &  0 &  0 &  1 &  1 &  0 &  0 &  0 \\
      0 &  1 &  1 &  0 &  0 &  1 &  0 &  0 \\
      0 &  0 &  1 &  0 &  0 &  0 &  1 &  0 \\
      0 &  0 &  0 &  1 &  0 &  0 &  0 &  1 \\
\hline
      0 &  0 &  0 &  0 &  1 &  0 &  0 &  0 \\
      0 &  0 &  0 &  0 &  0 &  1 &  0 &  0 \\
      \end{array}\right)
\end{equation}
where the lines separate input, measuring and output vertices. The corresponding graph is shown in Fig.~\ref{step-2}. Clearly, the block matrix
\begin{equation}
{\Gamma_0}_{\{1,2|3,4,5,6\}}^{\{3,4,5,6|7,8\}}=\left(\begin{array}{cc|cccc}
 1 &  0 &  0 &  1 &  1 &  0  \\
      0 &  1 &  1 &  0 &  0 &  1  \\
      0 &  0 &  1 &  0 &  0 &  0  \\
      0 &  0 &  0 &  1 &  0 &  0  \\
\hline
      0 &  0 &  0 &  0 &  1 &  0  \\
      0 &  0 &  0 &  0 &  0 &  1  \\
      \end{array}\right)
\end{equation}
has maximal rank which implies that $\Gamma_0$ is a basic graph. The measuring vertex $3$ fulfills the condition (ii). It has no self-link but it is connected to the measuring vertex $4$. Thus the set $\{3,4\}$ is removable and we can build the Schur complement
\begin{equation}
\Gamma_1:=\8X_{\{3,4\}}\Gamma_0=\left(\begin{array}{cc|cc|cc}
0& 1& 0& 1& 0& 0\\
1& 0& 1& 0& 0& 0\\
\hline
0& 1& 0& 1& 1& 0\\
1& 0& 1& 0& 0& 1\\
\hline
0& 0& 1& 0& 0& 0\\
0& 0& 0& 1& 0& 0\\
\end{array}\right)
\end{equation}
which is a graph on the remaining vertices $\{1,2|5,6|7,8\}$. This operation is illustrated by Fig.~\ref{step-2}, Step~1.

Now, the measuring vertex $5$ satisfies condition (ii). It has no self-link but it is connected to the other remaining measuring vertex $5$. The set $\{5,6\}$ is therefore removable with respect to the graph $\Gamma_1$ and we can build the Schur complement as well
\begin{equation}
\Gamma_2:=\8X_{\{5,6\}}\Gamma_1=\left(\begin{array}{cc|cc}
0&0&1&0\\
0&0&0&1\\
\hline
1&0&0&1\\
0&1&1&0
\end{array}\right)
\end{equation}
which yields the final graph $\Gamma_2$ that only lives on the input vertices $\{1,2\}$ and the output vertices $\{7,8\}$ (Fig.~\ref{step-2}, Step 2).

\begin{figure}[h]
\vspace*{13pt}
\centerline{\epsfig{file=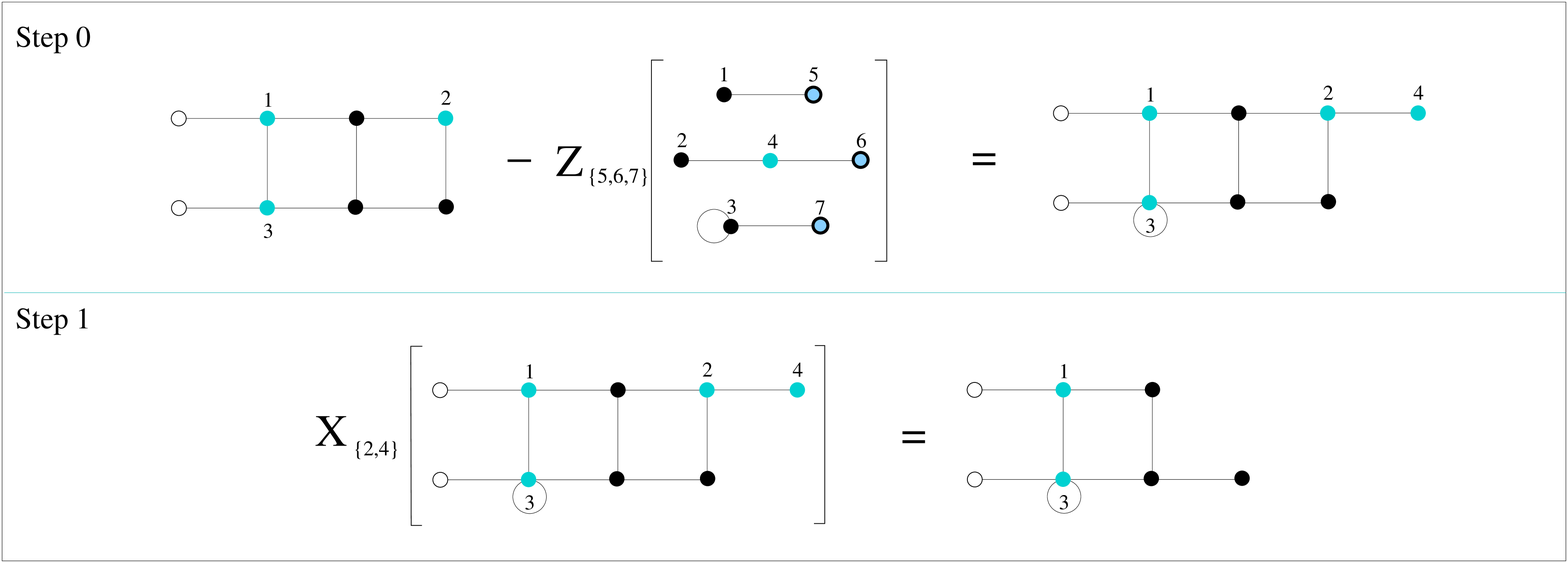, width=13cm}} 
\vspace*{13pt}
\fcaption{Step~0: The left figure shows a graph $\Gamma$ with measuring vertices $\{1,2,3\}$ representing the underlying multipartite entanglement recource for performing algorithms. 
The right figure represents a measuring graph $\Lambda$ which corresponds to the measurement strategy we wish to perform at the outputs $\{1,2,3\}$. $\{4\}$ labels the auxiliary vertex and syndrome vertices are labeled by $\{5,6,7\}$. We perform local measurements in $x$-basis at $\{1\}$, in $z$-basis at $\{2\}$ and in $y$-basis at $\{3\}$. This yields, combined with the corresponding measurement graph, the graph $\Gamma_0$. Step~1: The measuring vertices $\{2,4\}$ of $\Gamma_0$ are removable. According to (ii) we obtain the graph $\Gamma_1$.}
\label{step-00}
\end{figure}

\paragraph*{\it Example II:}
Fig~\ref{step-00} shows a basic graph $\Gamma$ which represents the underlying entanglement resource for running the algorithm under consideration. For illustration, we choose here an algorithm which is encoded by measuring the qubit at position $\{1\}$ in $x$-basis, the qubit localized at $\{2\}$ in $z$-basis, and the qubit sitting at $\{3\}$ in $y$-basis. This scheme corresponds to a measurement graph $\Lambda$ which is also depicted in Fig.~\ref{step-00}.    
\begin{figure}[h]
\vspace*{13pt}
\centerline{\epsfig{file=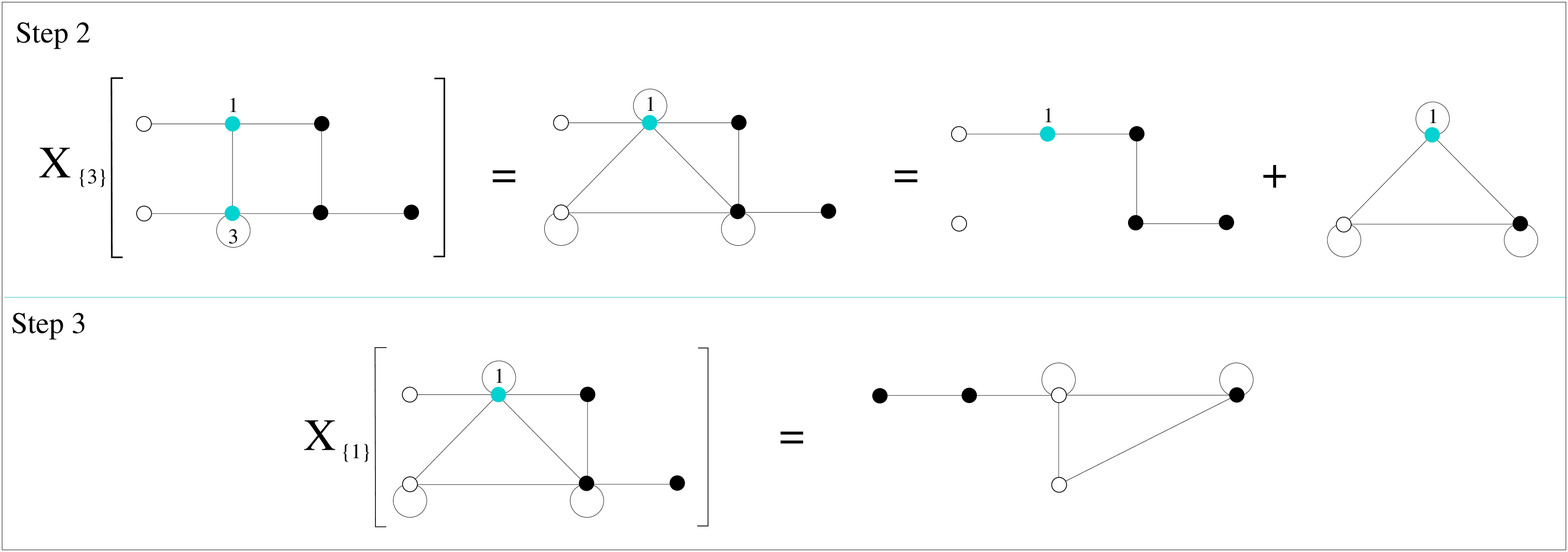, width=13cm}} 
\vspace*{13pt}
\fcaption{Step~2: The measuring vertices $\{3\}$ of $\Gamma_1$ has a self-link and can be removed by (i) which yields the graph $\Gamma_2$. Step~3: The measuring vertices $\{1\}$ of $\Gamma_2$ has a self-link and can also be removed by (i) which finally yields the graph $\Gamma_3$.}
\label{step-22}
\end{figure}

According to Theorem~\ref{x-measure}, we can substitute the graphs $\Gamma$ and $\Lambda$ by one graph $\Gamma_0=\Gamma-\8Z_{\{5,6,7\}}\Lambda$ which represents an equivalent entanglement recource on which only measurement operations in $x$-basis has to be performed (Fig.~\ref{step-00}, Step 0). First we remove, as shown by Fig.~\ref{step-00}, Step 1, the vertices $\{2,4\}$ by making use of (ii) which yields the graph $\Gamma_1$. This graph differs from the underlying graph $\Gamma$ in the following manner: The measuring vertex $\{3\}$ has an additional self-link and the vertex $\{2\}$ is removed from the cluster. In view of that we can formulate the following role:
\begin{itemize}
\item
A measuring vertex, which is measured in $x$-basis, remains unchanged. 
\item
A measuring vertex, which is measured in $z$-basis, is removed from the cluster. 
\item 
A measuring vertex, which is measured in $y$-basis, gets an additional self-link and is measured in $x$-basis.
\end{itemize}
The measurement vertices $\{1,3\}$ of the graph $\Gamma_1$ can completely be removed by applying the strategy, which is described above. The vertex $\{3\}$ has a self-link and can immediately by removed by (i) as shown by Fig.~\ref{step-22}, Step~2. Again we obtain a graph $\Gamma_2$ with one remaining measuring vertex $\{1\}$ that has a self-link. Thus we can apply the strategy (i) once more which produces the final graph $\Gamma_3$ (Fig.~\ref{step-22}, Step~3)

The present example shows one interesting feature which goes along with the removing of measuring vertices. The underlying graph $\Gamma$ represents a configuration of qubits that are arranged in a cubic lattice with next-neighbor Ising interaction. Measuring qubits at positions $\{1,2,3\}$ in $x$-, $z$-, and $y$-basis respectively is equivalent to preparing a configuration of qubits, according to the graph $\Gamma_3$, that evolve via this pattern of two-qubit Ising interactions and self-interactions. This can be interpreted as simulating the interaction, given by $\Gamma_3$, by using next-neighbor Ising interactions and by performing local measurement operations. 
 
\section{Conclusion}
We close the present paper by discussing an idea what the ``essence" of an algorithm is. The procedure, given in Section~\ref{algos} by a finite number of steps $\>00\>01\to n1\>$, describes how an algorithm can be implemented by local measurement operations on cluster states. In the first step, the information is encoded into the cluster via preparing the input qudits at positions $I$ in a state that carries the information we wish to process. The remaining qudits are individually prepared in the standard state. Within the next step we apply one elementary step of the dynamics which is given by a weighted graph $\Gamma$ on the union of the set $I$ of input vertices, the set $O$ of output vertices, and mutually disjoint sets of measuring vertices $\{M_t|t=0,\cdots,n\}$. 

In the first measurement round, we measure the input qudits in $x$-basis and we perform a local graph measurement on the qudits which are localized at $M=M_0$. The corresponding measurement graph $\Lambda$ lives on the union of the set $M=M_0$, a set $K$ of auxiliary vertices, and a set $L$ of syndrome vertices. After this procedure has been performed, we obtain a random measurement result $(p^I,q^L)$ (Fig.~\ref{alg-graph}).  

By making use of Theorem~\ref{x-measure}, only measurements in $x$-basis are needed in the first measurement round: We measure the qudits which are localized at the input vertices $I$ and the measuring vertices $K M$ in $x$-basis. This produces a random measurement outcome $p^{I K M}$. Provided $\Gamma-\8Z_L\Lambda$ is a basic graph with input vertices $I$ output vertices $J$ and measuring vertices $M K$, Theorem \ref{thm-byprod} can be applied which states that each random measurement outcome $p^{I K M}$ can be compensated by applying a phase space translation that operates at the positions $J$. This translation corresponds to the vector $(a^J|b^J)\in\Xi^J$ which is explicitly be given by  
\begin{eqnarray}
a^J=\Gamma^J_{J M}\overline{(\Gamma-\8Z_L\Lambda)}^{J M}_{I K M}p^{I K M}
&\text{ and }& b^J=\overline{(\Gamma-\8Z_L\Lambda)}^{J}_{I K M}p^{I K M} \; .
\end{eqnarray}
Here $\overline{(\Gamma-\8Z_L\Lambda)}^{J M}_{I K M}$ is a right inverse for the block matrix $(\Gamma-\8Z_L\Lambda)_{J M}^{I K M}$. 

\begin{figure}[h]
\vspace*{13pt}
\centerline{\epsfig{file=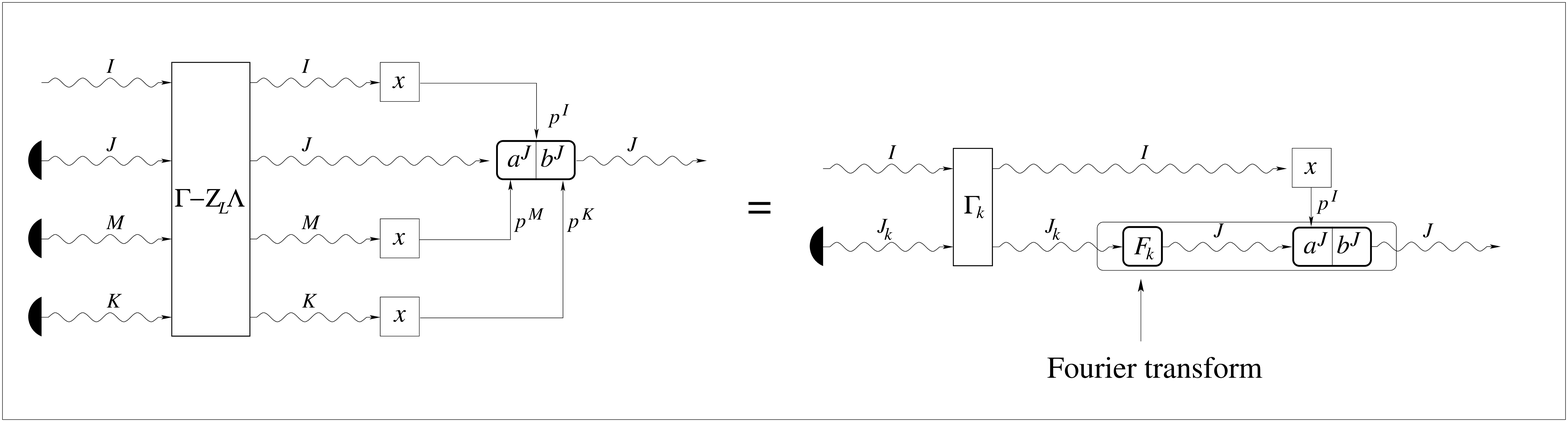, width=13cm}} 
\vspace*{13pt}
\fcaption{Removing measuring vertices: Prepare the qudits at $J K M$ individually in the standard state, apply an elementary step of the dynamic associated with $\Gamma-\8Z_L\Lambda$, measure the qudits at $I K M$ in $x$-basis. Then perform, conditioned by the measurement outcome, a local unitary operation at $J$ that compensates the measurement result. This operation coincides with a operational scheme that does not make use of graph measurements: Prepare the qudits at $J_k$ individually in the standard state, apply an elementary step of the dynamic associated with $\Gamma_k$, measure the input qudits at $I$ in $x$-basis, finally perform, conditioned by the measurement outcome, a local unitary operation at $J$ that compensates the measurement result.}
\label{alg-4}
\end{figure}

With help of Theorem~\ref{thm-remove} it follows that the measuring vertices $K M$ can be removed successively from the graph $\Gamma-\8Z_L\Lambda$ which we choose as initial graph $\Gamma_0=\Gamma-\8Z_L\Lambda$ with input vertices $I$, output vertices $J_0=O M_1\cdots M_n$ and measuring vertices $K_0=M K$. After a finite number $k$ of steps we obtain a graph $\Gamma_k$ that just have input vertices $I$, output vertices $J_k$ and a local Fourier transform $F_k$ that operates on the qudit positions $J_k\setminus (J_0\cap J_k)$. Thus the channel that is represented by Fig.~\ref{alg-graph} can be realized by the simplified scheme which is represented by Fig.~\ref{alg-4}. The local Fourier transform $F_k$ can be absorbed into the compensating part and it is the ``essence of the algorithm" what is left: The graph $\Gamma_k$ with input vertices $I$, output vertices $O$ and measuring vetices $M_1,\cdots, M_n$. However, the graph $\Gamma_k$ is not unique and depends on in which order the measuring vertices are removed.
 
\nonumsection{Acknowledgements}
\noindent
I am very grateful to Hans Briegel and Robert Raussendorf for supporting this investigation with many ideas. I would also like to acknowledge Reinhard Werner for his support and interesting and helpful discussions. 
Funding by the European Union project EQUIP (contract IST-1999-11053) is gratefully acknowledged. This research project is also supported by the Deutsche Forschungsgemeinschaft (DFG-Schwerpunkt "Quanteninformationsverarbeitung").
 


\appendix{\\ Stabilizer codes and algebras}
In Section~\ref{stab-codes} we discuss the general theory of stabilizer codes. From the mathematical perspective the basic object is the abelian C*-algebra, the stabilizer algebra, $\6A(I,J,K|\Lambda)$, which is naturally represented on the Hilbert space $\H{J}$. This part of the appendix serves several useful lemmas that enable to perform an explicit decomposition of the natural representation of $\6A(I,J,K|\Lambda)$ into irreducibles.    

The basic objects, which are concerned here are the operators
\begin{equation}\label{encode-app}
\1v_{[\Lambda|q^L]}=\sqrt{d}^{|JK|} \ \Phi_{[{I K L}]}^*{u}(\Lambda)\Phi_{[{J K}]}\Pi_{[q^L]}
\end{equation} 
which play a central role for obtaining an explicit decomposition of the natural representation. In addition to that we are also concerned with the closely related operators     
\begin{equation}
{\1u}_{[\Gamma|p^{I K}]}=\sqrt{d}^{|IK|} \Phi_{[p^{I K}]}^*{u}(\Gamma)\Phi_{[{K J}]}
\end{equation}
which describe the operation which has been performed after the measurement in $x$- basis produces the random result $p^{I K}$. The operator ${\1u}_{[\Gamma|{I K}]}={\1u}_{[\Gamma|0^{I K}]}$ corresponds to the standard measurement outcome. It is an isometry whenever $\Gamma\in\8{BG}(I,J;K)$ is a basic graph, i.e. the block matrix $\Gamma_{I K}^{J K}$ is injective: 

\begin{lem}\label{tech-lem-00}
If and only if  $\Gamma$ is a basic graph in $\8{BG}(I,J;K)$, then ${\1u}_{[\Gamma|{I K}]}$ is an isometry from $\H{I}$ into $\H{J}$. 
\end{lem}
\proof{
In order to decide whether ${\1u}_{[\Gamma|{I K}]}$ is an isometry, we now study the operator ${\1u}_{[\Gamma|{I K}]}^*{\1u}_{[\Gamma|{I K}]}$. Its action on a function  $\psi\in\H{I}$ is given according to 
\begin{equation}\label{iso-test}
({\1u}_{[\Gamma|{I K}]}^*{\1u}_{[\Gamma|{I K}]}\psi)(q_2^I)
=d^{|IK|}\int\8dq_1^I \ \1K(q_2^I|q_1^I) \psi(q_1^I) 
\end{equation}
where the ``integral kernel" $\1K$ is given by 
\begin{equation}\label{equ-step-1}
\1K(q_2^I|q_1^I)=\int\8dq_1^K \8dq_2^K \8dq^J \ \tau(-\Gamma|q_2^{I K}+q^J)\tau(\Gamma|q_1^{IK}+q^J)
\end{equation}
The map $\tau_\Gamma:q\mapsto\tau(\Gamma|q)$ is a project representation of the additive group $\7F^{I J K}_d$ which satisfies the identity $\tau(\Gamma|q_1+q_2)=\tau(\Gamma|q_1)\tau(\Gamma|q_2)\chi(q_1|\Gamma q_2)$. Making use of this fact, equation (\ref{equ-step-1}) turns into the following form:
\begin{eqnarray}\label{equ-step-2}
\1K(q_2^I|q_1^I)&=&\tau(-\Gamma|q_2^I)\tau(\Gamma|q_1^I)\int\8dq_1^K\8dq_2^K \tau(-\Gamma|q_2^K)\tau(\Gamma|q_1^K) 
\nonumber\\
&\times&\chi(\Gamma^K_Iq_2^I|-q_2^K) \chi(\Gamma^K_Iq_1^I|q_1^K) 
\int\8dq^J \chi(q^J|\Gamma^J_{I K}(q_1^{IK}-q_2^{IK})) \; .
\end{eqnarray}
The integration over the variables $q^J$ can explicitly been performed where we keep in mind that the Fourier transform on a finite abelian group yields  $\int\8dq^J\chi(p^J|q^J)=\delta(p^J)$. Thus we conclude that the integral kernel $\1K$ fulfills 
\begin{eqnarray}\label{equ-step-3}
\1K(q_2^I|q_1^I)&=&\tau(-\Gamma|q_2^I)\tau(\Gamma|q_1^I)
\int\8dq_1^K\8dq_2^K \tau(-\Gamma|q_2^K)\tau(\Gamma|q_1^K) 
\nonumber\\
&\times&\chi(\Gamma^K_Iq_2^I|-q_2^K) \chi(\Gamma^K_Iq_1^I|q_1^K)\delta(\Gamma^J_{IK}(q_1^{IK}-q_2^{IK})) \; .
\end{eqnarray}
We exploit the invariance of the Haar measure $\8dq_2^K$ and perform the variable transformation $q_3^K=q_2^K-q_1^K$ which yields
\begin{eqnarray}\label{equ-step-4}
\1K(q_2^I|q_1^I)&=&\tau(-\Gamma|q_2^I)\tau(\Gamma|q_1^I)
\int\8dq_1^K\8dq_3^K \ \tau(-\Gamma|q_3^K)\chi(\Gamma^K_Iq_2^I|-q_3^K)
\nonumber\\
&\times&\chi(\Gamma^K_I(q_1^I-q_2^I)-\Gamma^K_Kq^K_3|q_1^K) \delta(\Gamma^J_{I}(q_1^{I}-q_2^{I})-\Gamma^J_Kq_3^K) 
\end{eqnarray}
where we have used the identity 
$\tau(-\Gamma|q_3^K+q_1^K)\tau(\Gamma|q_1^K)=\tau(-\Gamma|q_3^K)\chi(\Gamma^K_Kq^K_3|-q_1^K)$.
Now we perform the integration over the variables $q_1^K$ and we find that the equation 
\begin{eqnarray}\label{equ-step-5}
\1K(q_2^I|q_1^I)=\tau(-\Gamma|q_2^I)\tau(\Gamma|q_1^I)\int\8dq_3^K \ \tau(-\Gamma|q_3^K)\chi(\Gamma^K_Iq_2^I|-q_3^K)\delta(\Gamma^{JK}_{IK}(q_1^I-q_2^I-q^K_3)) 
\end{eqnarray}
is valid. If we assume that $\Gamma$ is a basic graph, then $\Gamma^{J K}_{I K}(q_1^I-q_2^I-q^K_3)=0$ implies that $q_1^I=q_2^I$ and $q_3^K=0$ hold. Inserting this into (\ref{equ-step-5}) yields via performing the integration over $q^K_3$: 
\begin{eqnarray}\label{equ-step-6}
\1K(q_2^I|q_1^I)&=&d^{-|K|}\delta(q_1^I-q_2^I) \; . 
\end{eqnarray}
Finally, we insert this identity into Equation (\ref{iso-test}) and we obtain 
\begin{equation}
({\1u}_{[\Gamma|{IK}]}^*{\1u}_{[\Gamma|{IK}]}\psi)(q_2^I)=d^{|I|}\int\8dq_1^I \ \delta(q_1^I-q_2^I) \psi(q_1^I) 
=\psi(q_2^I)
\end{equation}
which implies that ${\1u}_{[\Gamma|{I K}]}$ is an isometry. On the other hand, one sees from equation (\ref{equ-step-5}) that the function $\1K$ cannot be supported on the diagonal $q_1^I=q_2^I$ if there exists a non-zero vector $q^{IK}$ with $\Gamma_{IK}^{JK}q^{IK}=0$. Hence, if $\Gamma$ is not basic, then ${\1u}_{[\Gamma|{IK}]}$ cannot be an isometry}

The following lemma presents a useful relation between the isometries $\1v_{[\Lambda|q^L]}$ and the operators $\1u_{[\Gamma|p^I]}$.

\begin{lem}\label{tech-lem-0}
Let $\Gamma$ be a graph with input vertices $I$, output vertices $J M$ and let $\Lambda$ be a graph without input vertices, output vertices $M$ and measuring vertices $K$. Then the identity 
\begin{equation}
\1v_{[\Lambda|q^L]}^*{\1u}_{[\Gamma|p^I]}=\Pi_{[q^L]}^*\1u_{[\Gamma-\Lambda|0^{MK},p^I]}
\end{equation}  
holds for all register configurations $q^L,p^I$.
\end{lem}
\proof{
We insert the definition of the operators $\1v_{[\Lambda|q^L]}$ and ${\1u}_{[\Gamma|p^I]}$ and we compute the expression
\begin{equation}
\1v_{[\Lambda|q^L]}^*{\1u}_{[\Gamma|p^I]}=\sqrt{d}^{|IKL|}
\Pi_{[q^L]}^*\Phi^*_{[{KM}]}u(-\Lambda)\Phi_{[{KL}]}\Phi^*_{[p^I]}u(\Gamma)\Phi_{[{JM}]} \; .
\end{equation}
The operator $\1v_{[\Lambda|q^L]}$ does not operate on the qudits that are localized at input vertices $I$. As a consequence, the co-isometry $\Phi^*_{[p^I]}$ commutes with $\1v_{[\Lambda|q^L]}$. Moreover, the operator $u(\Gamma)$ operates trivially at the positions $K L$ and commutes therefore with $\Phi_{[{K L}]}$. Keeping in mind that $|M|=|L|$, this yields  
\begin{eqnarray}
\1v_{[\Lambda|q^L]}^*{\1u}_{[\Gamma|p^I]}=\sqrt{d}^{|IKM|}
\Pi_{[q^L]}^*\Phi^*_{[0^{KM},p^I]}u(\Gamma-\Lambda)\Phi_{[{JKLM}]}\Pi_{[q^L]}^*
\1u_{[\Gamma-\Lambda|0^{KM},p^I]}
\end{eqnarray} 
where we have used the fact that $u(\Gamma-\Lambda)=u(\Gamma)u(-\Lambda)$ holds which is a direct consequence of the definition (\ref{defi-dyn})}

\begin{lem}\label{tech-lem-1}
For all classical register configurations $q^L$, the linear operator $\1v_{[\Lambda|q^L]}$ is an isometry which maps the input Hilbert space $L_2(\7F_d^I)$ into the output Hilbert space $L_2(\7F_d^J)$.
\end{lem}
\proof{
The operator $\Pi_{[q^L]}$ is an isometry and it is therefore sufficient to show that the operator 
\begin{equation}\label{encode-app}
\1u_{[\Lambda|IKL]}=\sqrt{d}^{|JK|} \ \Phi_{[{IKL}]}^*{u}(\Lambda)\Phi_{[{JK}]}
\end{equation}
is isometric. We may interprete the vertices $I L$ as ``inputs" and $K$ as ``measuring" vertices. According to the assumption (G2), the block matrix $\Lambda^{J K}_{I K L}$ is invertible, in other words, $\Lambda$ is a basic graph in $\8{BG}(I L,J;K)$. Thus Lemma \ref{tech-lem-00} can be applied which proves that $\1v_{[\Lambda|q^L]}=\1u_{[\Lambda|I K L]}\Pi_{[q^L]}$ is an isometry}

Before the next lemma is presented, we note here a useful observation which will be used several times within this appendix. For any subset $N\subset M$ the vector ${\xi}_{[N]}$ is invariant under all shifts $\1x(q^N)$ and ${\zeta}_{[N]}$ is invariant under all multiplier $\1z(p^N)$. Moreover, according to the definition of $x$- and $z$- basis, the identities $\1x(q^N){\zeta}_{[N]} ={\zeta}_{[q^N]}$ and $\1z(p^N)\xi_{[N]}=\xi_{[p^N]}$ are valid. This implies the useful relations
\begin{eqnarray}\label{pi-rel}
&&\1w(p^M|q^M)\Pi_{[N]}=\Pi_{[q^N]}\1w(p^{M\setminus N}|q^{M\setminus N})
\\\label{phi-rel}
&&\1w(p^M|q^M)\Phi_{[N]}=\Phi_{[q^N]}\1w(p^{M\setminus N}|q^{M\setminus N}) \; .
\end{eqnarray}

\begin{lem}\label{tech-lem-2}
The range of the isometry $\1v_{[\Lambda|q^L]}$ is the multiplicity space of the character $\varpi_{[q^L]}$ of the stabilizer algebra $\6A(I,J,K|\Lambda)$ which is uniquely determined by its the expectation values on Weyl operators according to
\begin{equation}\label{char-ew}
\Ew{\varpi_{[q^L]}}{\1w(\Lambda^{J}_{J K}q^{J K}|q^{J})}=\tau(\Lambda|q^{J K L}) \; ,
\end{equation}
where  $(\Lambda^{J}_{J K}q^{J K}|q^{J})$ is in the isotropic subspace $S_{[I,J,K|\Lambda]}$.
\end{lem}
\proof{
For proving the eigenvalue equation (\ref{stab-ew}), we consider a register configuration $q^{J K}$ which belongs to the kernel of the block matrix $\Lambda^{I K}_{J K}$ . With help of (\ref{pi-rel}) and (\ref{phi-rel}) we verify that the identity 
\begin{equation}
c \ \1v_{[\Lambda|q^L]}=\Phi_{[{I K L}]}^*{u}(\Lambda)\1x(q^{J K L})\Phi_{[{J K}]}\Pi_{[L]}
\end{equation} 
holds, where $c$ is an appropriate normalization constant. In the second step, we make use of the 
Weyl commutation relations as well as the fact that ${u}(\Lambda)$ implements the symplectic transformation $(p^M|q^M) \to (p^M-\Lambda q^M|q^M)$ with $M=I J K L$. This gives
\begin{multline}
c \ \1v_{[\Lambda|q^L]}
=\tau(-\Lambda|q^{J K})\1w(\Lambda^{J}_{J K}q^{J K}|q^{J})
\Phi_{[-\Lambda^{I K}_{J K}q^{J K},0^L]}^*{u}(\Lambda)
\1z(\Lambda^L_{J K}q^{J K})
\\
\times\1x(q^L)\Phi_{[{J K}]}\Pi_{[L]} \; .
\end{multline}
We exchange the order of $\1z(\Lambda^L_{J K}q^{J K})$ and $\1x(q^L)$, making use of the assumption that $q^{J K}$ belongs to the kernel of $\Lambda^{I K}_{J K}$ and applying the relation (\ref{pi-rel}) once more which yields the identity:  
\begin{equation}
c \ \1v_{[\Lambda|q^L]}
=\tau(-\Lambda|q^{J K})\chi(-\Lambda^L_{J K}q^{J K}|q^L)\1w(\Lambda^{J}_{J K}q^{J K}|q^{J})\Phi_{[{I K L}]}^*
{u}(\Lambda)\Phi_{[{J K}]}\Pi_{[q^L]} \; .
\end{equation}
According to the definition of the isometry $\1v_{[\Lambda|q^L]}$, we find   
\begin{equation}
\1v_{[\Lambda|q^L]}
=\tau(-\Lambda|q^{J K})\chi(-\Lambda^L_{J K}q^{J K}|q^L)
\1w(\Lambda^{J}_{J K}q^{J K}|q^{J K})\1v_{[\Lambda|q^L]}\; .
\end{equation}
Since there are no edges between syndrome vertices we conclude $\tau(\Lambda|q^{J K})\chi(\Lambda^L_{J K}q^{J K}|q^L)$ coincides already with $\tau(\Lambda|q^{J K L})$
which proves the lemma}

\begin{lem}\label{tech-lem-3}
The family $\{\1v_{[\Lambda|q^L]}|q^L\in\7F_d^L\}$ of isometries is mutually orthogonal
\begin{equation}\label{orth}
\1v_{[\Lambda|q_1^L]}^*\1v_{[\Lambda|q_2^L]}=\delta(q_1^L-q_2^L) \ \11_I
\end{equation}
and complete
\begin{equation}\label{compl}
\sum_{q^L\in\7F_d^L}\1v_{[\Lambda|q^L]}\1v_{[\Lambda|q^L]}^*=\11_J \; .
\end{equation}
\end{lem}
\proof{
We compare the expectation values of two characters $\varpi_{[p_1^L]}$ and $\varpi_{[p_2^L]}$ for a given Weyl operator $\1w(\Lambda^{J}_{J K}q^{J K}|q^{J})$ which belongs to the stabilizer algebra (see (\ref{char-ew})) by analyzing the ratio of both eigenvalues
\begin{eqnarray}\label{frac-ew}
\frac{\tau(\Lambda|q^{J K}+q^L_1)}{\tau(\Lambda|q^{J K}+q_2^L)}
=\frac{\tau(\Lambda|q^L_1)}{\tau(\Lambda|q_2^L)} \chi(\Lambda_L^{J K}(q^L_1-q_2^L)|q^{J K})
=\chi(\Lambda_L^{J K}(q^L_1-q_2^L)|q^{J K}) \; .
\end{eqnarray}
Here we have used the fact that $\tau(\Lambda|q_2^L)=\tau(\Lambda|q_1^L)=1$ since there are no edges
which connect syndrome vertices. The fraction of the eigenvalues (\ref{frac-ew}) is $=1$ for all configurations $q^{J K}\in\8{ker}\Lambda_{J K}^{I K}$ only if $\Lambda_L^{J K}(q^L_1-q_2^L)$ lies in the range of $\Lambda^{J K}_{I K}$, i.e. there is a configuration
$q^{I K}$ such that
\begin{equation}
\Lambda_L^{J K}(q^L_1-q_2^L)+\Lambda^{J K}_{I K}q^{I K}=0
\end{equation}
holds. By assumption (G2), the matrix $\Lambda^{J K}_{I K L}$ is invertible which implies $q^{I K}=0$ and $q^L_1=q_2^L$. Thus the characters $\varpi_{[q_1^L]}$ and $\varpi_{[q_2^L]}$ coincide only if $q_1^L=q_2^L$. By Lemma \ref{tech-lem-2}, the ranges of the isometries  $\1v_{[\Lambda|q_1^L]}$ and $\1v_{[\Lambda|q_2^L]}$ are therefore orthogonal to each other if $q_1^L$ and $q_2^L$ are distinct.

The range of each isometry $\1v_{[\Lambda|q^L]}$ has dimension $d^{|I|}$. We have already shown the orthogonality (\ref{orth}) which implies that the projection, defined by the left-hand side of  (\ref{compl}), projects onto a $d^{|L|+|I|}$ dimensional space. By assumption (G1) we have $|I|+|L|=|J|$ and (\ref{compl}) follows}

\appendix{\\ Operations on cluster states}
In this part of the appendix we develop the tools for tackle local measurement operations in $x$-basis (this implicitly covers also measurements in $y$- and $z$- bases) which are applied to cluster states in an appropriate manner. A Weyl operator, which acts on the input Hilbert space $\H{I}$ commutes with the isometry ${\1u}_{[\Gamma|{I K}]}$ according to the following useful relation:

\begin{lem}\label{tech-lem-4}
Let $\Gamma$ be a graph with input vertices $I$, output vertices $J$ and measuring vertices $K$. Then for each register configuration $p^I$ the commutation relation 
\begin{equation}\label{com-rell}
\tau(\Gamma|q^{I J K}) \ {\1u}_{[\Gamma|{I K}]}\ \1w(p^I|q^I) 
=\1w(\Gamma^J_{I J K}q^{I J K}|q^J) \ {\1u}_{[\Gamma|{I K}]}
\end{equation}
holds if $q^{J K}$ is a solves the equations
\begin{eqnarray}
&&p^I=\Gamma^{I}_{I J K}q^{I J K} 
\\
&&0^{K}=\Gamma^{K}_{I J K}q^{I J K} \; .
\end{eqnarray}
\end{lem}
\proof{
We first recall the fact that the isometry $\Phi_{[{J K}]}$ is invariant under all shift operators $\1x(q^{J K})$ (see (\ref{phi-rel})) and that the unitary operator ${u}(\Gamma)$ implements the symplectic transformation $(p^I|q^{I J K})\mapsto(p^I-\Gamma q^{I J K}|q^{I J K})$, i.e. the corresponding commutation relation (\ref{com-rel-0}) holds. From this we conclude that 
\begin{equation}\label{eq-abc}
\tau(\Gamma|q^{I J K}) \ {u}(\Gamma)\Phi_{[{J K}]} \ \1w(p^I|q^I)=\1w(p^I-\Gamma q^{I J K}|q^{I J K}){u}(\Gamma)\Phi_{[{J K}]}
\end{equation}
is valid. Now we apply the co-isometry $\Phi_{[{I K}]}^*$ on both sides of the Equation (\ref{eq-abc}) which yields
\begin{equation}
\tau(\Gamma|q^{IJK}){\1u}_{[\Gamma|{I K}]} \ \1w(p^I|q^I)
=\1w(-\Gamma^J_{IJK}q^{IJK}|q^J){\1u}_{[\Gamma|\Gamma^{I K}_{I J K}q^{I J K}-p^I]} \; .
\end{equation}
Here we have used the identity (\ref{phi-rel}). Therefore the commutation relation (\ref{com-rell}) follows if $p^{I}-\Gamma^{I K}_{I J K}q^{I J K}=0$ is fulfilled}

A similar result is given by the next lemma which can be used to prove the existence of compensating maps (see Subsection \ref{subsec-comp} for the definition of this notion). 

\begin{lem}\label{tech-lem-4-a}
Let $\Gamma$ be a graph with input vertices $I$, output vertices $J$ and measuring vetices $K$, then the identity  
\begin{equation}\label{comp-app}
{\1u}_{[\Gamma|{I K}]}
=\tau(\Gamma|q^{K J})\1w(-\Gamma^{J}_{K J}q^{K J}|q^{J})^*{\1u}_{[\Gamma|p^{I K}]}
\; .
\end{equation}
holds if the equation
\begin{equation}\label{eq-compensate}
\Gamma^{I K}_{K J}q^{K J}+p^{I K}=0
\end{equation}
is fulfilled. 
\end{lem}
\proof{
Due to the shift invariance of the isometry $\Phi_{[{K J}]}$, i.e. (\ref{phi-rel}), the identity 
\begin{equation}
d^{|IK|} {\1u}_{[\Gamma|p^{I K}]}
=\Phi_{[p^{I K}]}^*{u}(\Gamma)\1x(q^{K J})\Phi_{[{K J}]}
\end{equation} 
is fulfilled for all classical register configurations $q^{K J}$. We exchange the order of the operators ${u}(\Gamma)$ and $\1x(q^{K J})$ which yields by keeping (\ref{phi-rel}) once more in mind:
\begin{eqnarray}
d^{|IK|}  {\1u}_{[\Gamma|p^{I K}]}
=
\tau(-\Gamma|q^{K J})\1w(-\Gamma^{J}_{K J}q^{K J}|q^{J})
\Phi_{[p^{I K}+\Gamma^{I K}_{K J}q^{K J}]}^*{u}(\Gamma)\Phi_{[{K J}]} \; .
\end{eqnarray}
This implies that (\ref{comp-app}) is true if the classical register configurations $p^{I K}$ and $q^{K J}$ solve the equation $\Gamma^{I K}_{K J}q^{K J}+p^{I K}=0$}

A heuristic analysis of standard measurement operations in $x$-basis suggests that they are directly related to opeartions on graphs. The following two lemmas serves as key ingredients for proving that this intuition is indeed correct. 

\begin{lem}\label{tech-lem-5}
Let $\Gamma\in\8{BG}(I,J;K)$ be a basic graph and let $N\subset K$ be removable. Then the identity 
\begin{equation}\label{coincide-ii}
\Gamma q^{I J K}=(\8X_N\Gamma)q^{I J M}
\end{equation}
holds, $M=K\setminus N$, for all $q^{J K}$ which solve the equation 
\begin{equation}\label{cond-ii}
\Gamma^{I K}_{I J K}q^{I J K}=p^I
\end{equation}
for some $p^I,q^I\in\7F_d^I$. In particular, it follows that 
\begin{equation}\label{coincide-iii}
\tau(\Gamma|q^{I J K})=\tau(\8X_N\Gamma|q^{I J M})
\end{equation}    
is true for all $q^{I J K}$ that fulfill {\rm (\ref{cond-ii})}.
\end{lem}
\proof{
We first split the system of equations (\ref{cond-ii}) into two parts 
\begin{eqnarray}
&&\Gamma^{I M}_{I J M}q^{I J M}+\Gamma^{I M}_Nq^N=p^I
\\
&&\Gamma^{N}_Nq^N+\Gamma^N_{I J M}q^{I J M}=0
\end{eqnarray}
which is equivalent to 
\begin{eqnarray}
&&\Gamma^{I M}_{I J M}q^{I J M}+\Gamma^{I M}_Nq^N=p^I
\\\label{subst-ii}
&&q^N=\bar\Gamma^N_N\Gamma^N_{I J M}q^{I J M}
\end{eqnarray}
since, by making use of the assumption that $N$ is removable, $\Gamma^N_N$ has an inverse $\bar\Gamma^N_N$. Thus we can substitute $q^N$ into the first equation  which yields 
\begin{eqnarray}
(\8X_N\Gamma)^{I M}_{I J M}q^{I J M}=
\Gamma^{I M}_{I J M}q^{I J M}-\Gamma^{I M}_N\bar\Gamma^N_N\Gamma^N_{I J M}q^{I J M}
=p^I \; .
\end{eqnarray}
Finally one verifies that (\ref{coincide-ii}) holds by making use of(\ref{subst-ii}) via inserting it into the left hand side of (\ref{coincide-ii}). Equation (\ref{coincide-iii}) follows immediately from  the definition of the phases $\tau(\Gamma|\cdot)$}

\begin{lem}\label{tech-lem-6}
Let $\Gamma\in\8{BG}(I,J;K)$ be a basic graph and let $N\subset K$ be removable. Then the operator 
\begin{equation}
{\1u}_{[\Gamma|I K]}^*{\1u}_{[\8X_N\Gamma|I K\setminus N]}\in\7C\11_I
\end{equation}
is a multiple of the identity in $\6A(I)$.
\end{lem}
\proof{
By Lemma \ref{tech-lem-4} the isometries ${\1u}_{[\Gamma|I K]}$ and ${\1u}_{[\8X_N\Gamma|I M]}$, $M=K\setminus N$, commute with a Weyl operator $\1w(p^I|q^I)$ according to (\ref{com-rell}). Since $\Gamma$ is a basic graph, the block matrix $\Gamma^{I K}_{J K}$ is surjective and for each $p^I$ there is $q^{I J K}$ which fulfills the equations $p^I-\Gamma^I_Iq^I=\Gamma^I_{J K}q^{J K}$ and $\Gamma^K_{J K}q^{J K}=-\Gamma_I^Kq^I$. This implies that the equations $p^I=(\8X_N\Gamma)^I_{I J M}q^{I J M}$ and $(\8X_N\Gamma)^M_{I J M}q^{I J M}=0$ are valid due to the fact that $\Gamma^N_N$ is invertible. Hence we conclude that the relation  
\begin{equation}\label{equ-1-1}
\tau(\Gamma|q^{I J K}) \ {\1u}_{[\Gamma|{I K}]}\ \1w(p^I|q^I)=\1w(-\Gamma^J_{I J K}q^{I J K}|q^J) \ {\1u}_{[\Gamma|{I K}]}
\end{equation}
as well as
\begin{equation}\label{equ-1-2}
\tau(-\8X_N\Gamma|q^{I J M}) \ \1w(p^I|q^I)^* \ {\1u}_{[\8X_N\Gamma|{I M}]}^*\ 
={\1u}_{[\8X_N\Gamma|{I M}]}^* \ \1w(-(\8X_N\Gamma)^J_{I J M}q^{I J M}|q^J)^* 
\end{equation}
is fulfilled. Due to Lemma \ref{tech-lem-5} we also conclude that the identities $\tau(\8X_N\Gamma|q^{I J M})=\tau(\Gamma|q^{I J K})$ and $\1w((\8X_N\Gamma)^{J}_{I J M}q^{J M}|q^{J})=\1w(\Gamma^{J}_{I J K}q^{I J K}|q^{J})$ are also true. Combining both equations (\ref{equ-1-1}) and (\ref{equ-1-2}) implies that  
\begin{eqnarray}
{\1u}_{[\8X_N\Gamma|I M]}^*{\1u}_{[\Gamma|I K]} \ \1w(p^I|q^I)=\1w(p^I|q^I)\ {\1u}_{[\8X_N\Gamma|I M]}^*{\1u}_{[\Gamma|I K]} 
\end{eqnarray}
holds, i.e. ${\1u}_{[\8X_N\Gamma|IM]}^*{\1u}_{[\Gamma|IK]}$ commutes with all Weyl operators. Thus, it is indeed a multiple of the identity}

A further important relation shows that the application of a Fourier transform to a cluster state can also be interpreted as an operation on the underlying graph by adding additional vertices and edges.

\begin{lem}\label{tech-lem-7}
Let $\Gamma\in\8{BG}(I,J N;K)$ be a basic graph with output vertices $N J$, $J\cap N=\emptyset$. Moreover let $\nu\mathpunct:N\mapsto M$ be a bijective map and $\widehat\nu$ the connecting graph that connects each vertex $n\in N$ with its counterpart $\nu(n)\in M$. Then $\Gamma+\widehat\nu$ is a basic graph in $\8{BG}(I,J M;K N)$ and the identity 
\begin{equation}
F_{[{\widehat\nu}_N^M]}{\1u}_{[\Gamma|IK]}={\1u}_{[\Gamma+\widehat\nu|IKN]}
\end{equation}
holds, where $F_{[{\widehat\nu}_N^M]}$ is the local Fourier transform associated with the connecting graph $\widehat\nu$. 
\end{lem}
\proof{
According to our assumption $\Gamma$ is a basic graph to which we add the connecting graph $\widehat\nu$ which yields the graph $\Gamma+\widehat\nu$ with input vertices $I$, output vertices $J M$ and measuring vertices $K N$. This is indeed a basic graph since the kernel of the block matrix $(\Gamma+\widehat\nu)^{J M N K}_{I N K}=\Gamma_{I N K}^{I J N}+\widehat{\nu}_N^M$ is trivial. This can be observed as follows: Suppose that 
$(\Gamma+\widehat\nu)^{JMNK}_{INK}q^{INK}=0$ holds. Then this implies $\widehat\nu_N^Mq^N=0$ and therefore $q^N=0$.  Thus we conclude that $\Gamma_{I N K}^{I J N}q^{I N K}=\Gamma_{I K}^{I J N}q^{I K}=0$ is valid. But since $\Gamma$ is basic, we find $q^{I K}=0$. 

Applying the isometry ${\1u}_{[\Gamma|I K]}$ to a vector $\psi\in\H{I}$ can explicitly be computed according to 
\begin{equation}
({\1u}_{[\Gamma|I K]}\psi)(q^{J N})=c  \int\8dq^{I K} \tau(\Gamma|q^{I J N  K})\psi(q^I)
\end{equation}
where the constant $c=\sqrt{d}^{-|I|-|K|}$ is for normalization. Concatenating this operator with the Fourier transform  $F_{[{\widehat\nu}_N^M]}$ for the connecting graph $\widehat\nu$ yields  
\begin{multline}\label{iso-fourier}
(F_{[{\widehat\nu}_N^M]}{\1u}_{[\Gamma|I K]}\psi)(q^{J M}) = c' \int\8dq^{N}\ \chi(q^M|\widehat\nu^M_Nq^N) \ ({\1u}_{[\Gamma|I K]}\psi)(q^{J N})
\\
=
c c' \int\8dq^{INK}\ \chi(q^M|\widehat\nu^M_Nq^N)\tau(\Gamma|q^{IJNK})\psi(q^I)
\end{multline}
with $c'=\sqrt{d}^{-|N|}$. Making use of the definition of the bi-character $\chi(\cdot|\cdot)$ and the phases $\tau(\cdot|\cdot)$ we verify that the identity
\begin{multline}
\chi(q^M|\widehat\nu^M_Nq^N)\tau(\Gamma|q^{I J N  K})
=
\exp\biggl[\frac{\pi\8i}{d}\<q^{I J N  K},\Gamma q^{I J N  K}\>\biggr] \exp\biggl[{\frac{2\pi\8i}{d}\<q^M,\widehat\nu^M_Nq^N\>}\biggr]
\\
=
\exp\biggl[\frac{\pi\8i}{d} \<q,(\Gamma+\widehat\nu) q\>\biggr]=
\tau(\Gamma+\widehat\nu|q)
\end{multline}
holds for all register configurations $q=q^{IJKMN}$. By inserting this relation into the Equation (\ref{iso-fourier}) and by using the definition of the isometry ${\1u}_{[\Gamma+\widehat\nu|I K N]}$ gives rise to the desired identity  
\begin{equation}\label{iso-fourier}
(F_{[{\widehat\nu}_N^M]}{\1u}_{[\Gamma|I K]}\psi)(q^{J M})=({\1u}_{[\Gamma+\widehat\nu|I K N]}\psi)(q^{J M})
\end{equation}
which completes the proof of the lemma}
\end{document}